\documentclass[aip,jcp,amsmath,amssymb,reprint,]{revtex4-1}

\usepackage{graphicx}
\usepackage{dcolumn}
\usepackage{bm}

\usepackage[utf8]{inputenc}
\usepackage[T1]{fontenc}
\usepackage{mathptmx}

\begin{document}

\title{Interface between graphene and liquid Cu from molecular dynamics simulations}

\author{Juan Santiago Cingolani}
\affiliation{Chair for Theoretical Chemistry and Catalysis Research Center, Technische Universit{\"a}t M{\"u}nchen, Lichtenbergstr. 4, 85747 Garching, Germany}
\author{Martin Deimel}
\affiliation{Chair for Theoretical Chemistry and Catalysis Research Center, Technische Universit{\"a}t M{\"u}nchen, Lichtenbergstr. 4, 85747 Garching, Germany}
\author{Simone K{\"o}cher}
\thanks{Current address: School of Physics / CRANN, Trinity College Dublin, 42A Pearse St., Dublin 2, Ireland}
\affiliation{Chair for Theoretical Chemistry and Catalysis Research Center, Technische Universit{\"a}t M{\"u}nchen, Lichtenbergstr. 4, 85747 Garching, Germany}
\author{Christoph Scheurer}
\affiliation{Chair for Theoretical Chemistry and Catalysis Research Center, Technische Universit{\"a}t M{\"u}nchen, Lichtenbergstr. 4, 85747 Garching, Germany}
\author{Karsten Reuter}
\affiliation{Chair for Theoretical Chemistry and Catalysis Research Center, Technische Universit{\"a}t M{\"u}nchen, Lichtenbergstr. 4, 85747 Garching, Germany}
\affiliation{Fritz-Haber-Institut der Max-Planck-Gesellschaft, Faradayweg 4-6, 14195 Berlin, Germany}
\author{Mie Andersen}
\email{mie.andersen@ch.tum.de}
\affiliation{Chair for Theoretical Chemistry and Catalysis Research Center, Technische Universit{\"a}t M{\"u}nchen, Lichtenbergstr. 4, 85747 Garching, Germany}

\begin{abstract}
Controllable synthesis of defect-free graphene is crucial for applications since the properties of graphene are highly sensitive to any deviations from the crystalline lattice. We focus here on the emerging use of liquid Cu catalysts, which has high potential for fast and efficient industrial-scale production of high-quality graphene. The interface between graphene and liquid Cu is studied using force field and \textit{ab initio} molecular dynamics, revealing a complete or partial embedding of finite-sized flakes. By analyzing flakes of different sizes we find that the size-dependence of the embedding can be rationalized based on the energy cost of embedding versus bending the graphene flake. The embedding itself is driven by the formation of covalent bonds between the under-coordinated edge C atoms and the liquid Cu surface, which is accompanied by a significant charge transfer. In contrast, the central flake atoms are located around or slightly above 3~$\text{\AA}$ from the liquid Cu surface and exhibit weak vdW-bonding and much lower charge transfer. The structural and electronic properties of the embedded state revealed in our work provides the atomic-scale information needed to develop effective models to explain the special growth observed in experiments where various interesting phenomena such as flake self-assembly and rotational alignment, high growth speeds and low defect densities in the final graphene product have been observed.
\end{abstract}

\maketitle

\section{Introduction}\label{intro}
Liquid Cu has recently emerged as an interesting catalyst for the synthesis of high-quality single-layer graphene through chemical vapor deposition (CVD) carried out at temperatures just above the melting point of Cu.\cite{Geng2012,Wu2012,Wu2013,Geng2014,Zeng2016,Xue2019,Zheng2019} For solid Cu, advanced catalyst preparation techniques such as temperature-gradient-driven annealing of Cu foils to produce the desired Cu(111) facet \cite{Xu2017} and chemical–mechanical polishing of the foil,\cite{Nguyen2015} have recently been used to demonstrate seamless stitching of growing rotationally aligned neighboring graphene flakes, yielding low defect densities in the final macroscopic sheet. However, achieving the perfectly flat and defect-free solid Cu catalyst required for such growth remains difficult, and graphene grown on single-crystal Cu(111) surfaces thus often exhibits a high degree of grain boundaries and other defects.\cite{Gao2010} Liquid Cu, in contrast, is an easy route to a smooth catalyst surface void of defects and grain boundaries. In general, this translates into fewer defects in the synthesized graphene sheet compared to corresponding CVD at solid Cu,\cite{Li2009,Bhaviripudi2010,Vlassiouk2011,Kim2012,Li2016,Huet2017} which is of utmost importance for applications of the final graphene product in e.g.\ electronics or optics, since its otherwise remarkable properties in these application areas can be heavily influenced by defects.\cite{Novoselov2005,Allen2010} Significantly enhanced graphene growth rates can also be achieved on liquid Cu.\cite{Zheng2019} Together with the prospect of shearing and directly separating graphene from the molten Cu surface,\cite{Saedi2020} thus avoiding transfer-induced defects and allowing for continuous production, this makes the liquid catalysis method particularly interesting for industrial applications. From a more fundamental point of view, a peculiarity of the graphene / liquid Cu interface is the self-assembly and rotational alignment of graphene flakes, which has been observed to extend over micrometer scales \cite{Geng2012,Wu2012,Wu2013} and whose origin has been heavily debated.\cite{Geng2014,Zeng2016,Xue2019,SciencePaper}

In spite of the promising experimental results, theoretical studies of the graphene / liquid Cu interface are scarce. Li {\em et al.}\ carried out molecular dynamics (MD) simulations based on density-functional tight binding to investigate the growth process beginning from C dimers.\cite{Li2014} They suggested that the liquid Cu could play a role in healing defects arising in the graphene sheet during growth. Larger-scale MD simulations are possible using variable charge reactive force fields like COMB3.\cite{Liang2012,Liang2013} COMB3 has recently been applied to the graphene / liquid Cu interface in simulations aimed at investigating the Cu melting and cooling processes \cite{Klaver2015} and a possible role of the gas flow in the CVD reactor during growth for the rotational alignment of graphene flakes.\cite{Xue2019} Another work based on a self-developed C-Cu empirical force field focused on semi-molten Cu(111) surfaces \cite{Xu2020} and aimed at explaining the rotational alignment and seamless stitching observed in the above-mentioned experiments carried out with polished Cu(111) foils.\cite{Nguyen2015,Xu2017}

In this contribution we focus on the liquid Cu catalyst, as we believe this represents the most promising catalyst for efficient, large-scale production of single-crystal graphene.\cite{SciencePaper} Based on MD simulations carried out with the COMB3 force field, we find that graphene flakes embed into the liquid Cu surface, which is an interesting growth motif that has also previously been discussed for the solid and semi-molten Cu(111) surface.\cite{Yuan2014,Xu2020} We further carry out large-scale \textit{ab initio} MD (AIMD) simulations using a C$_{54}$ graphene flake and a simulation box containing 1589 Cu atoms, allowing us to verify the force field predictions without compromising a realistic description of the liquid Cu surface. By investigating graphene flakes of different sizes as well as the full monolayer sheet, we reveal the size-dependent structural and electronic characteristics of the embedded state. This allows us to rationalize the embedding in terms of competing bending strain and embedding energies. In particular, we find that the embedded state enables additional covalent bonding concentrated around the edges of the flake, whereas the central part of the flake is characterized by weak vdW-bonding at an adsorption height around or slightly above 3~$\text{\AA}$. At the atomic scale, the liquid Cu surface itself, however, is relatively unperturbed by the presence of the graphene sheet. Indeed, calculations of the radial distribution function (RDF) rule out a possible local ordering or crystallization of the liquid Cu under the flake. Our detailed structural and electronic characterization of the size-dependent flake embedding is a first starting point for understanding various experimentally observed phenomena, e.g.\ the self-assembly and rotational alignment of flakes and the special growth characterized by high growth speeds and low defect densities.

\section{Computational details}\label{comp}

\subsection{Force field MD}\label{FFMD}
The force field MD (FFMD) simulations were conducted using slab models of a Cu surface with graphene adsorbed on one side of the slab. Periodic boundary conditions were used in the dimensions parallel to the surface. Multiple independent simulations were run using hexagonal graphene flakes with zigzag edges of size C$_{54}$, C$_{150}$, C$_{294}$, C$_{384}$ and C$_{600}$. These simulations employed a 67.043~$\text{\AA}$ $\times$ 67.043~$\text{\AA}$ simulation box containing 9360 Cu atoms. None of the Cu atoms were kept frozen.
After equilibration the thickness of the employed Cu slab was $\sim$27~$\text{\AA}$ and the smallest distance between any two C atoms in the periodic images of the largest C$_{600}$ flake was about 20~$\text{\AA}$. Furthermore, a separate simulation investigated a full graphene sheet covering the entire Cu surface.
This simulation employed a 31.974~$\text{\AA}$ $\times$ 29.820~$\text{\AA}$ simulation box containing 1728 Cu atoms and 364 C atoms. After equilibration the thickness of the Cu slab was $\sim$23.4~$\text{\AA}$. The size of the simulation box corresponds to a (13$\times$7) cell employing the rectangular 4-atom graphene unit cell of (1$\times \sqrt{3}$) and the optimum COMB3 graphene lattice constant of 2.46~$\text{\AA}$, and was chosen to minimize stress in the graphene lattice. A previous literature study employing AIMD simulations has shown that the lattice constant of free-standing graphene is almost constant (to within 0.005~$\text{\AA}$) in the temperature range of 0-2000~K,\cite{Pozzo2011} thus we do not expect that the temperatures employed in the present study cause any significant stress in the graphene lattice either. All simulations were carried out with the variable charge reactive force field COMB3 \cite{Liang2012,Liang2013} as implemented in the LAMMPS code \cite{Plimpton1995} and using a 1~fs timestep.

The following procedure was employed to melt and equilibrate the system in the NVT ensemble. Beginning from a solid Cu slab, the system was initialized with velocities drawn from a Maxwell-Boltzmann distribution corresponding to 1500~K and quickly melted by applying a Langevin thermostat with a characteristic timescale of 100~fs for 5~ps at the same temperature, which is substantially above both the experimental melting temperature of Cu (1358 K) and the melting temperature predicted by the COMB3 force field (1140-1145~K \cite{Klaver2015}). Then, to aid in achieving equilibration of the system, a Berendsen thermostat with a characteristic timescale of 1000~fs was run for an extra 5~ps at 1370~K. At this point the graphene flake was added on top of the molten slab ensuring that no C atom was closer than 2.8~$\text{\AA}$ to any Cu atom. This initial configuration of the flake was fixed and the Cu system was consecutively attached to the Langevin and Berendsen thermostats described above, but setting the temperature to 1370~K in both cases. As before, the system was evolved for 5~ps with each thermostat. During these simulation steps the flake can become embedded in the liquid Cu; independent runs where no constraints were applied to the C atoms verified that this configuration is not a consequence of fixing the flake in space during the initial equilibration. The system containing a full graphene sheet was prepared by separately applying the 10~ps initialization and equilibration procedure to a Cu slab and a free-standing graphene sheet, then bringing the two systems in contact ensuring that no C atom was closer than 1.8~$\text{\AA}$ to any Cu atom
and evolving the system as above with a Langevin thermostat at 1380~K for 2~ps as well as a Berendsen thermostat at 1370~K for 5~ps without applying any constraints to the C atoms.

All productions runs were then carried out using a Nos\'{e}-Hoover thermostat with a characteristic timescale of 50 fs and a chain length of one at 1370 K, which is a typical temperature for liquid Cu CVD growth.
At least the first 5~ps were used for additional equilibration and ignored during property evaluation. All thermostats were used as implemented in the ASE code.\cite{Hjorth_Larsen_2017}

\subsection{DFT and AIMD}
A smaller version of the system with the C$_{54}$ flake was also studied using AIMD. In particular, we employed a 28.881~$\text{\AA}$ $\times$ 28.881~$\text{\AA}$ $\times$ 63.168~$\text{\AA}$ simulation box containing 1589 Cu atoms. The thickness of the Cu slab after force field-based pre-equilibration is similar to the value reported above (24.3~$\text{\AA}$ in the present case), i.e.\ the reduction in system size is primarily in the lateral dimensions of the Cu slab. Periodic boundary conditions were used in all three dimensions and the vacuum separation between consecutive slabs (including the embedded graphene flake) is above 33~{\AA}.
These simulations were driven by the full-potential, all-electron density-functional theory (DFT) package FHI-aims \cite{Blum2009} using light default settings for the basis set and integration grids, the PBE exchange-correlation functional,\cite{Perdew1996} gamma point \textbf{k}-point sampling, and a dipole correction. Long-range vdW forces and the collective many-body response of the metallic substrate were considered through the effective pairwise-additive dispersion correction scheme vdW$^{\rm surf}$.\cite{Tkatchenko2009,Ruiz2012}

To minimize the amount of expensive \textit{ab initio} steps required for equilibration, the system was first simulated with the COMB3 force field: the same preparation as described in the previous section was carried out, followed by at least 10~ps under the Nos\'{e}-Hoover thermostat. After that the simulation was continued using the forces calculated by FHI-aims and under the same Nos\'{e}-Hoover thermostat at 1370 K. The simulated trajectory has a total length of 933 fs, of which the first 100 fs were used for equilibration and the remaining 833 fs for property evaluation.

Standard bulk and slab models were generated for comparison and validation purposes. Using the above settings in FHI-aims and a ($12\times12\times12$) \textbf{\textit{k}}-point grid we calculate an optimized solid Cu fcc lattice constant of 3.63~$\text{\AA}$, which is in good agreement with the experimental lattice constant of 3.61~$\text{\AA}$.\cite{Cu_lattice_constant} The adsorption height of graphene on solid Cu(111) was modeled in a periodic four-layered (1$\times$1) cell. To conform with previous studies,\cite{Olsen2011,Andersen2012,Andersen2019b} we fixed the lateral size of the cell using the Cu lattice constant (here we used the experimental lattice constant) and adapted the graphene lattice constant accordingly. We used FHI-aims and the settings described above, a (12$\times$12) \textbf{\textit{k}}-point grid, and a vacuum region of about 200~$\text{\AA}$. The bottom two Cu layers were fixed in their bulk-truncated position, while the upper two layers and the graphene lattice were relaxed until the maximum force on each atom fell below 0.05~eV/$\text{\AA}$. The calculated graphene adsorption height of 3.41~$\text{\AA}$ is in quite good agreement with previous studies that found values of 3.2--3.3~$\text{\AA}$.\cite{Olsen2011,Andersen2012,Andersen2019b}

\subsection{Force correlation}
In order to benchmark the forces predicted by the force field, we recalculated snapshots of FFMD simulations of the C$_{54}$ flake and the full monolayer graphene sheet with DFT. The flake employed a 19.254~$\text{\AA}$ $\times$ 19.254~$\text{\AA}$ $\times$ 30.876~$\text{\AA}$ simulation box containing 256 Cu atoms and the sheet a 12.298~$\text{\AA}$ $\times$ 12.780~$\text{\AA}$ $\times$ 29.141~$\text{\AA}$ simulation box with 120 Cu and 60 C atoms. For these DFT calculations we used the plane-wave code Quantum ESPRESSO \cite{Giannozzi2017} v.6.3 for the sheet and v.6.4.1 for the flake with the PBE exchange-correlation functional\cite{Perdew1996} and the semi-empirical D3 van der Waals (vdW) correction scheme by Grimme \textit{et al}.\cite{Grimme2010} The employed ultrasoft pseudopotentials were generated using the "atomic" code by A. Dal Corso  (v.5.0.2 svn rev. 9415) and a dipole correction was used.\cite{Bengtsson1999}
Along each (\textit{x},\textit{y}) cell direction, the Brillouin zone was sampled with an (n $\times$ m) grid of \textbf{\textit{k}}-points, where at least 31/\textit{a} \textbf{\textit{k}}-points were used, with \textit{a} being the cell length in $\text{\AA}$. Cutoffs for the orbitals (charge density) were 500~eV (5000~eV), respectively.
The force correlation shown in Supplementary Fig.\ S1 shows a quite good correlation, especially for the full graphene sheet. For the flake there is a small systematic deviation of the forces and certain large outliers are observed, which is caused by under-coordinated C atoms at the edge of the flake that are harder to describe with the force field.

\section{Results and discussion}\label{res}

We begin by studying a smaller graphene flake consisting of 54 C atoms. Here we consider a flake where the edge atoms are not passivated by hydrogen, but by the Cu surface. This choice is motivated by our recent \textit{ab initio} thermodynamics study \cite{Andersen2019b} where we found that hydrogen- and metal-passivated edges have very similar formation free energies under typical CVD growth conditions and that hydrogen-passivated flakes have very low adsorption energies as a result of weak van der Waals interactions with the Cu surface. Even if the formation of hydrogen-passivated flakes is thus possible, they would rapidly desorb at liquid Cu CVD temperatures and we therefore do not consider such flakes in the present study.

\begin{figure}
  \centering
  {%
  \setlength{\fboxsep}{0pt}%
  \setlength{\fboxrule}{1pt}%
  \fbox{\includegraphics[width=0.5\columnwidth]{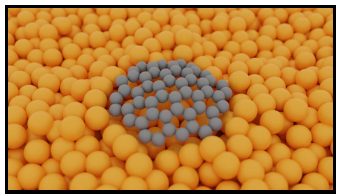}}%
  }%
  \caption{Snapshot from the FFMD trajectory of the C$_{54}$ flake showing the embedded state of the flake.}
  \label{fig:flake-md-snapshot}
\end{figure}

When adding the C$_{54}$ flake to the liquid Cu surface and equilibrating the system as described in the previous section, a peculiar structural motif appears. In essence, the flake embeds into the liquid surface as seen in the 3D representation of a snapshot from the FFMD trajectory (Fig.\ \ref{fig:flake-md-snapshot}). The under-coordinated edge C atoms thereby find themselves completely surrounded by the liquid Cu surface. This interesting structural motif calls for additional verification, as a benchmarking of the employed force field for graphene at liquid Cu has not been carried out in the previous literature studies of this system,\cite{Klaver2015,Xue2019} raising the question whether the embedded state could be an artifact of the force field. Embedding of a C$_{24}$ graphene flake has been verified by AIMD in the work of Ref.\ \cite{Xu2020} focusing on semi-molten Cu(111) surfaces. However, these AIMD simulations were carried out for a very small system size (3-layered ($6\times4$) Cu(111) slab), which is likely too small to appropriately capture the physical characteristics of a liquid surface. Here we therefore carry out large-scale AIMD for the C$_{54}$ flake using a realistic description of a liquid Cu surface with a simulation box containing 1589 Cu atoms, corresponding approximately to an 11-layered ($12\times12$) Cu(111) slab. Due to the large system size the total simulation time was limited to about 1~ps, of which the first 0.1~ps was used for pre-equilibration and thermostatting. Employing a running average over subsets of the trajectory, each subset containing 10 frames (i.e.\ 10 fs), we calculate the evolution of the embedding (i.e.\ the average distance between the flake and the liquid Cu surface) during the AIMD, see Supplementary Fig.\ S2. The result shows that the embedding remains fairly constant during the trajectory and that the fluctuating flake height of $-0.2\pm0.2$~{\AA} is about one order of magnitude smaller than the adsorption height of graphene on solid Cu(111) calculated with the exact same settings (about 3.4~$\text{\AA}$). This result provides further evidence for the embedded state of the flake. However, we stress that the limited trajectory length does not yet allow us to make definite, quantitative conclusions on this height and that there could be a long-term slow drift. Nevertheless, qualitatively, the data we have is fully consistent with the one derived for the embedded flakes at the force field level.

Before engaging in a more quantitative analysis of the structural properties of the embedded state, it is illustrative to compare it conceptually with literature structural graphene growth motives proposed for the solid Cu catalyst. Static zero-Kelvin DFT calculations have addressed larger C clusters on Cu(111). For flake-like clusters with more than 13 atoms incorporation of defects in the form of 5-membered rings was found to lead to a dome-like structural motif similar to a C buckyball.\cite{Zhong2016}
Nevertheless, the flakes still reside on top of the surface. Another work considered only the defect-free C$_{13}$ flake consisting of three neighboring 6-membered rings, but further carried out AIMD simulations at 1200~K.\cite{Didar2018} These simulations also suggested that the flake remains adsorbed on top of the Cu surface. However, a substantial roughening of the surface was observed with the formation of Cu adatoms, which then attach to the undercoordinated edge C atoms. In Ref.\ \cite{Xu2020} it was suggested that the Cu(111) surface might be semi-molten, and then also allowed for an embedded motif, at temperatures representative for solid Cu CVD of $\sim$1273~K (1000~$^{\circ}$C). This is in disagreement with large-scale COMB3 MD simulations of the melting process of graphene-covered Cu(111) though,\cite{Klaver2015} where it was found that the close-packed Cu(111) facet is very stable up till the bulk melting point of Cu and even shows superheating effects. Similar results were found in MD simulations focusing on the melting of Cu surfaces based on an effective medium theory (EMT) potential.\cite{Hakkinen1992} Also, in experimental works the Cu(111) surface was found to have a higher stability towards pre-melting than the other low-index facets \cite{Wang2015b} and has indeed been confirmed to remain solid up till the Cu bulk melting temperature.\cite{Stock1980} Although experimental solid Cu CVD growth works often employ Cu foils rather than bulk crystals, the used Cu foils still have thicknesses of 10--100~$\mu$m,\cite{Li2009,Nguyen2015} which is orders of magnitude thicker than Cu thin films (10--100 nm in thickness) where low-temperature melting has been reported.\cite{Gromov2007} Overall, this suggests that the embedded structural motif is unique for liquid Cu CVD.

\begin{figure*}
  \centering
  \includegraphics{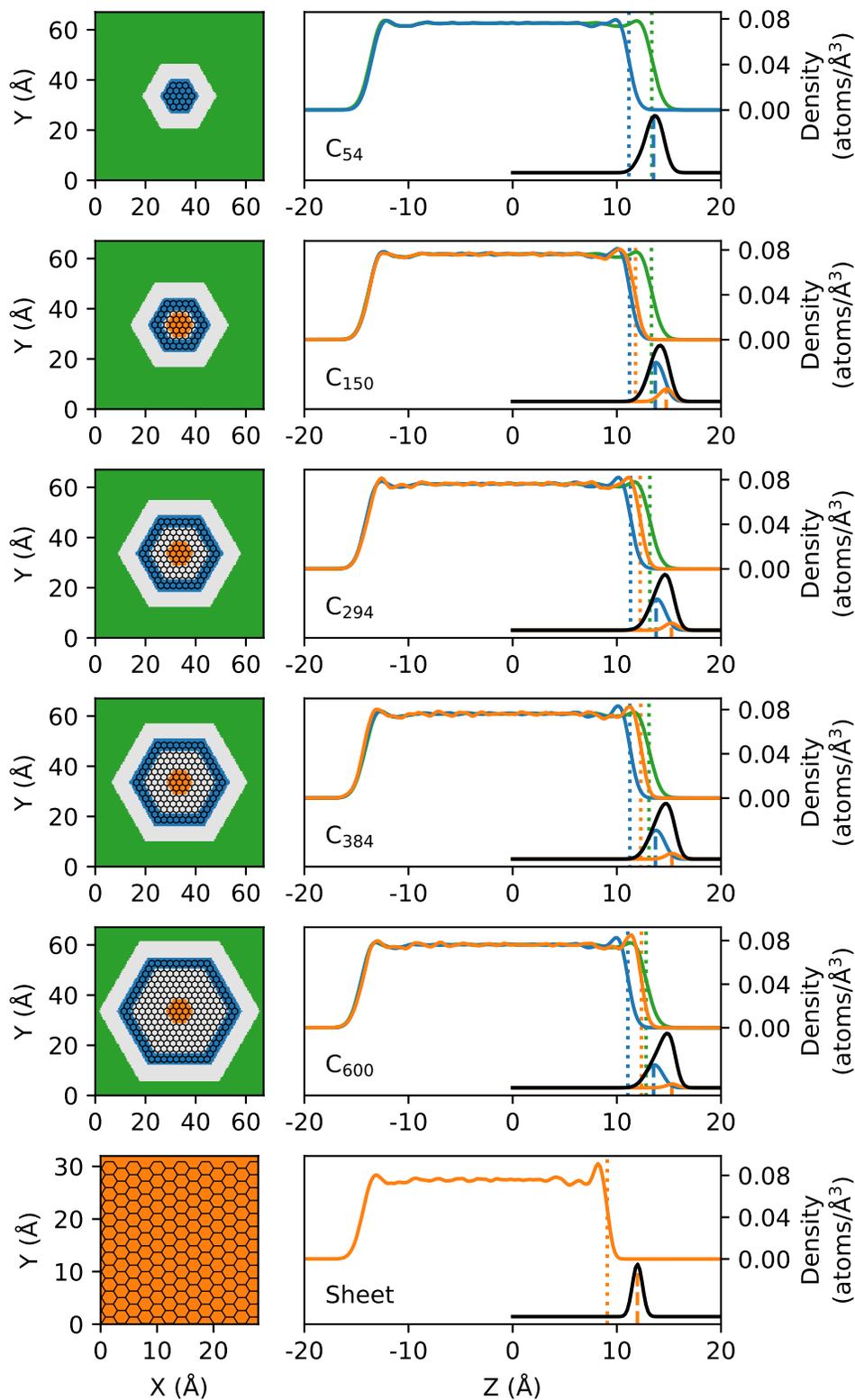}
  \caption{FFMD atom density profiles (right plots) in the direction perpendicular to the surface of systems with graphene flakes of different sizes (upper five plots) and a full graphene sheet (bottom plot). As shown in the left plots, the various contributions to the Cu atom density are differentiated, i.e.\ the uncovered Cu surface around the flake (green), the graphene-covered Cu surface near the flake edge (blue) and close to the center of the flake (orange). The dotted vertical lines indicate the position of the Cu surface (inflection point of the density). The C atom density (shown in black) has been normalized to arbitrary 2D density units and shifted vertically for clarity. Within the C curve, the vertical dashed lines highlight the average position of the C atoms belonging to edge (blue) and center (orange) regions (defined as for the Cu atoms). The results correspond to the averaging over independent trajectories adding up to at least 299.4~ps for each system.}
  \label{fig:z-profiles}
\end{figure*}

Having established the embedded structural motif for the smaller C$_{54}$ flake, we move on to quantify in more detail the structural properties of the embedded state and the dependence on the flake size through atom density profiles perpendicular to the surface obtained from the MD simulations (Fig.\ \ref{fig:z-profiles}). The results shown are for the FFMDs. A comparison of the AIMD and FFMD results for the smallest flake size is shown in Supplementary Fig.\ S3, demonstrating a quite good qualitative agreement. As a measure for the position of the Cu surface we use the inflection point of the Cu density profile (dotted vertical lines in Fig.\ \ref{fig:z-profiles}). The indentation in the liquid Cu surface is visible as the difference between the position of the uncovered Cu surface around the flake (green curve) and the position of the Cu surface below the flake (orange and blue curves). The average indentation of the Cu surface situated close to the edge of the graphene flake (blue curve) is relatively constant, ranging from 2.19~$\text{\AA}$ for the C$_{54}$ flake to 1.75~$\text{\AA}$ for the C$_{600}$ flake.
The smallest C$_{54}$ flake is completely embedded into the Cu surface, i.e.\ it is seen that the position of the flake (black curve) is very similar to the position of the free Cu surface (green curve). For the larger flakes a slight out-of-plane bending of the graphene flake is observed, i.e.\ the C atoms in the center of the flake (orange contribution to the C peak) are situated higher above the free Cu surface than those near the edge (blue contribution to the C peak), resulting in a skewing of the total C curve (black peak). See also the individual heights of all C atoms in the C$_{600}$ flake in Supplementary Fig.\ S4. This bending is also visible in the liquid Cu surface, where for larger flakes the position of the graphene-covered Cu surface near the center of the flake (orange curve) approaches the height of the uncovered Cu surface (green curve).

\begin{figure}
  \includegraphics[trim=0 20 0 14, clip]{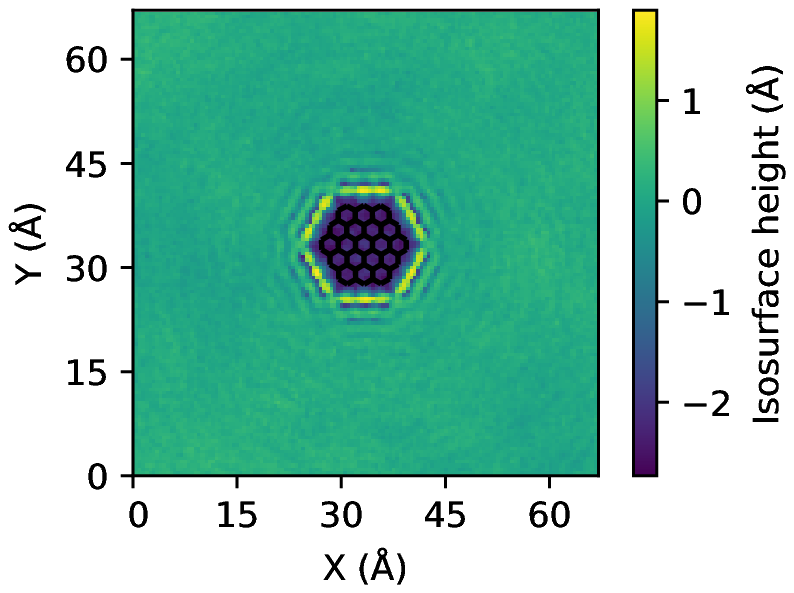}
  \includegraphics[trim=0 0 0 12, clip]{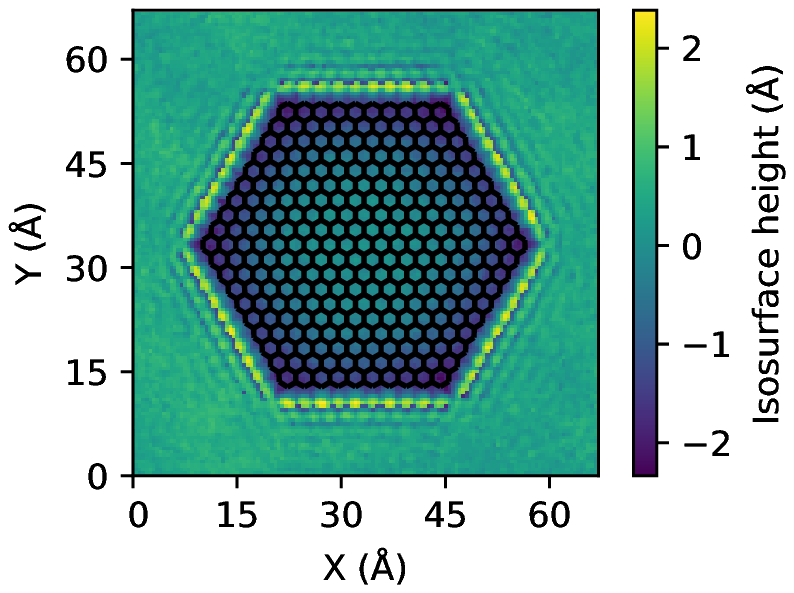}
  \caption{FFMD isodensity surfaces of the Cu atom density for the simulations with the C$_{54}$ flake (top) and the C$_{600}$ flake (bottom). Rotations of the graphene flakes during the MD trajectory around an axis perpendicular to the surface have been filtered out. The resulting position of the graphene flake is represented in black. The height zero is set to the average height of the isodensity surface far away from the flake. The results correspond to the averaging over trajectories of 998.8~ps for the C$_{54}$ flake and 584.9~ps for the C$_{600}$ flake.}
  \label{fig:2d-profiles}
\end{figure}

\begin{figure}
  \centering
  \includegraphics[trim=1 10 1 46, clip]{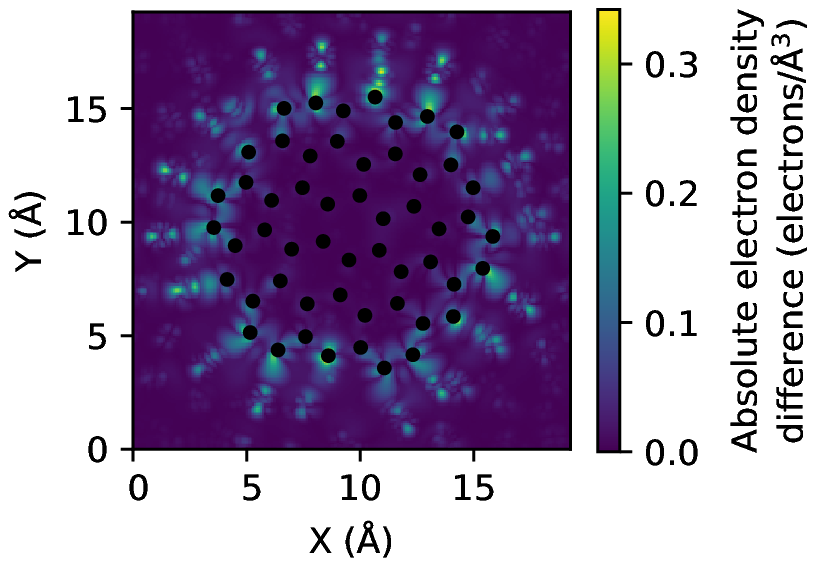}
  \caption{DFT electronic density difference plot (combined system minus isolated flake and Cu surface) calculated from a snapshot of the FFMD trajectory. The positions of the C atoms are shown in black. For any given ($X,Y$) point the maximum density difference along the $Z$-direction is shown. This allows to visually compare the charge transfer of individual C atoms located at slightly different $Z$ heights.}
  \label{fig:dens-diff}
\end{figure}

The structural effects of the Cu-graphene interactions can also be observed in Fig.\ \ref{fig:2d-profiles} where an isodensity surface of the Cu atom density at a value of 0.05 atoms/$\text{\AA}^3$ (i.e.\ close to the inflection point of the Cu density profile shown in Fig.\ \ref{fig:z-profiles}) manifests once again the emergence of a local indentation in the Cu surface to accommodate the graphene flake. More interestingly, a pattern is seen to form around the edges of the flake. Contrasting with the indentation below the flake, an increased height (about 2~$\text{\AA}$ above the remaining free Cu surface) is observed around the edges of the flake. The indentation and the induced patterning of the Cu isodensity surface can be rationalized by a strong interaction of the edge under-coordinated C atoms with the surface Cu atoms. This flake-surface interaction can be visualized by subtracting from the DFT-calculated electronic density of a snapshot of the FFMD trajectory of the combined system, the electronic densities of the isolated flake and Cu surface. Such a density difference plot is shown in Fig.~\ref{fig:dens-diff} and reveals that all charge transfer (and thereby stronger covalent bonding) is concentrated around the edge C atoms. For the central C atoms, however, the charge transfer is negligible and the bonding of these atoms to the Cu surface is therefore exclusively by weak van der Waals interactions.

The edge C atoms thus reduce their energy by embedding themselves into the liquid, i.e.\ these C atoms achieve a more favorable bonding configuration when they are surrounded by Cu atoms. For the smallest flakes, the edge atoms drag the rest of the flake along with them, resulting in a complete embedding. For the largest flakes a bending of the flake is induced, resulting in a complete embedding of only the edge atoms. We rationalize this finding in terms of competing embedding and strain energies. We expect the cost of embedding a C atom into the liquid Cu surface to be relatively constant for larger flake sizes. At the small vdW interaction of the overwhelming fraction of center C atoms, it corresponds essentially to the excluded volume cost for the Cu. As described above, the required indentation of the edge C atoms is relatively constant with the flake size (about 2~$\text{\AA}$).
Bending the flake as an alternative to spending the constant embedding cost requires then only a decreasing bending angle for larger flakes -- at concomitantly decreased strain energy per C atom. Above a critical flake size, this will thus favor bending over complete embedding.

We find that a graphene sheet at liquid Cu adsorbs at a height of 2.89~$\text{\AA}$ in the FFMD simulation employing a full graphene layer. At static solid Cu(111), vdW-corrected DFT functionals generally predict an adsorption height of graphene around 3.2--3.3~$\text{\AA}$.\cite{Olsen2011,Andersen2012,Andersen2019b} For comparison, the employed COMB3 force field predicts an adsorption height of 3.0~$\text{\AA}$ at solid Cu(111), which is thus slightly underestimated compared to the literature DFT results. Overall, this suggests that graphene at solid and liquid Cu adsorbs at very similar heights. In the force field simulations employing finite-sized graphene flakes, we find that the adsorption height in the central (orange-colored in Fig.\ \ref{fig:z-profiles}, not defined for the smallest C$_{54}$ flake) part of the flake is about 2.9--3.0~$\text{\AA}$, which is in good agreement with the sheet simulation and suggests that in flakes larger than C$_{150}$ the central region behaves similarly to the full graphene sheet.
In general, adsorption heights are very difficult to measure experimentally. At liquid Cu, an adsorption height of about 2~$\text{\AA}$ has recently been derived from \textit{in situ} X-ray reflectivity (XRR) measurements.\cite{SciencePaper} However, in these experiments the presence of C dissolved in the liquid Cu surface \cite{Zeng2014} could not be excluded, which was not taken into account in the assumed model on which the fitting of the XRR curves was based. This could have led to an underestimation of the graphene adsorption height.

As we have shown up till now, a finite-sized graphene flake at liquid Cu interacts strongly with the liquid surface through the under-coordinated edge atoms and is either completely or partially embedded into the liquid. In contrast, the interior of the flake behaves similarly to the full graphene sheet in terms of a vdW-dominated bonding to the substrate and a typical adsorption height around or slightly above 3~$\text{\AA}$. In the literature, a heavily debated question is the physics behind the experimentally observed micrometer-scale self-assembly and rotational alignment of graphene flakes at liquid Cu.\cite{Geng2012,Wu2012,Wu2013}
Various theoretical explanations for this behavior have been offered in recent literature,\cite{Geng2014,Zeng2016,Xue2019} however, unfortunately the theoretical results and explanations are generally not convincing. The role of gravitational forces suggested in Ref.\ \cite{Geng2014} can be directly ruled out due to the low atomic mass of C and the fact that in experimental setups the sample is often tilted out of a horizontal alignment.\cite{SciencePaper} In Ref.\ \cite{Zeng2016} the electrostatic potentials of smaller free-standing graphene flakes were calculated, and the fact that the potential was shown to have an anisotropic shape at a length scale of a few Angstroms away from the graphene edge was used as the argument why flakes self-align over micrometer length scales. While it seems plausible that electrostatic effects could play a role, a convincing theory needs to account for the more than four orders of magnitude difference in the length scales probed by the atomistic simulations and those observed in the experiments.
An alternative theoretical explanation has been offered in Ref.\ \cite{Xue2019} based on the flow of gases (mainly carrier gases such as argon or nitrogen) in the reactors used in experiments. In this work MD simulations of a graphene flake on liquid Cu subject to a Poiseuille flow of N$_2$ atoms parallel to the surface were carried out. However, the simulations are unfortunately completely unrealistic due to the high flow rates employed (to save computational cost). In effect, under the flow rates employed the graphene flakes drift along the Cu surface with a velocity of about 10 m/s (!), which is multiple orders of magnitude higher than flake velocities observed in experiments. And despite these unrealistically high flow rates, only a rather unconvincing just-above-noise rotational motion of the flakes towards alignment was observed in the MD simulations.

\begin{figure}
  \centering
  \includegraphics[width=\columnwidth]{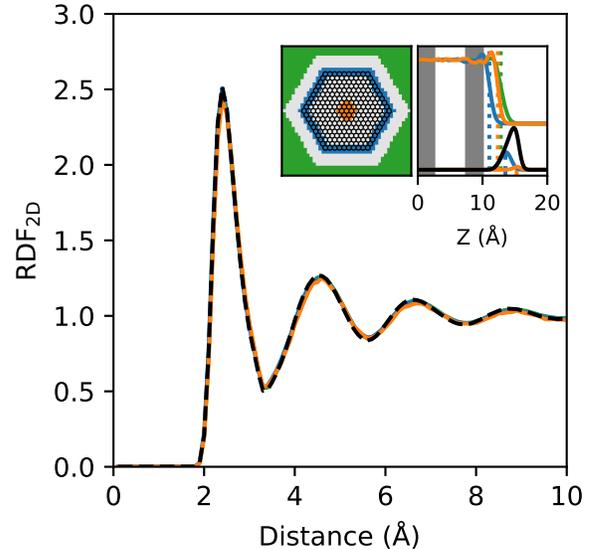}
  \caption{2D RDF calculated from the FFMD simulation of the C$_{600}$ flake within a box located in the middle of the Cu slab (from $Z=0~\text{\AA}$ to $Z=2.5~\text{\AA}$, dashed black curve) and in the subsurface region (from $Z=7.5~\text{\AA}$ to $Z=10~\text{\AA}$), in the latter case differentiating between Cu atoms located far away from the flake (green curve), below the flake near the edge C atoms (blue curve) or near the central C atoms (orange curve). The color coding of the different regions is consistent with Fig.\ \ref{fig:z-profiles} and reproduced in the left insert. Note that all curves are overlapping. The atom density profiles in the right inset are also reproduced from Fig.\ \ref{fig:z-profiles}. Here the shaded grey areas illustrate the $Z$ positions of the two boxes considered. The results correspond to the averaging over trajectories of 584.9~ps.}
  \label{fig:RDF}
\end{figure}

A further hypothesis could be that the self-alignment is mediated by the flake-surface interaction causing a local ordering or crystallization of the liquid Cu under the flake. For instance, recent COMB3-based MD simulations have shown that the presence of a graphene sheet on Cu(111) can slightly retard the surface melting.\cite{Klaver2015} In order to investigate this more generally for liquid Cu CVD, we calculate in Fig.\ \ref{fig:RDF} the two-dimensional (2D) RDF within a box (periodic in the $X$- and $Y$-directions, $Z$ height of 2.5 $\text{\AA}$) containing subsurface Cu atoms located in three different regions close to or far away from the flake. Surprisingly, the RDF curves from the three different regions (colored curves) are completely identical and also identical to an RDF calculated within a box located in the middle of the Cu slab far away from the surface (dashed black curve). All RDFs exhibit the characteristics of a liquid with broad peaks at shorter distances representing the coordination shells of the Cu atoms. A similar result is obtained when calculating the RDFs of the three regions within a box located at higher $Z$ coordinates, thus containing both surface Cu atoms and vacuum (see Supplementary Fig.\ S5). We attribute this lack of local ordering to the weak vdW-interactions with the substrate exhibited by all interior C atoms. All in all, these results do thus not support a local ordering or crystallization of the liquid Cu surface as driving force for the experimentally observed self-assembly.

Recently, based on a multiscale approach incorporating atomistic simulations into continuum modeling to rationalize novel \textit{in situ} experimental investigations, we have suggested that the experimentally observed self-assembly at the micrometer scale may instead be explained by short-range repulsive electrostatic interactions between the flakes and long-range attractive capillary interactions that are effective up till the capillary length of liquid Cu of 4~mm.\cite{SciencePaper} While this theory could well explain the experimental data, it only considered the large flake size limit where the interaction is dominated by the center flake atoms. As we show in this work, smaller flakes are either completely or partially embedded into the liquid and the flake-surface interaction is dominated by the edge flake atoms. This suggests that very different self-assembly results could be obtained if the growth was performed such that many small flakes are nucleated and they then coalesce while the flakes are still small. For the capillary interactions, the increased contribution of the edge atoms would likely make it necessary to develop more detailed models, going beyond the simple monopole-monopole interactions considered in our previous work. Also the electrostatic part of the flake-flake interaction was significantly simplified in our previous work. In the large flake limit, continuum modeling gave rise to a homogeneous charge distribution on the flakes, which is mirrored by the build-up of an image charge in the conducting surface. This allowed us to use simple electrostatic dipole-dipole interactions. As we show here from atomistic simulations, the charge transfer in reality happens primarily at the embedded edge flake atoms; see also Supplementary Fig.\ S6. The detailed charge distribution at hexagonal-shaped flakes presented here will allow to extend the electrostatic interaction to higher multipoles. We believe that the construction of such refined capillary-electrostatic models is the key to explain not only the length-scale of the self-assembly targeted in our previous work, but also the rotational alignment observed in experiments as well as the detailed dynamic behavior when flakes of different sizes and shapes are present at the surface simultaneously.\cite{SciencePaper} This is a topic of our ongoing work.

\section{Conclusions}
Based on FFMD and AIMD simulations employing realistic system sizes, we have shown that graphene flakes at liquid Cu embed into the liquid surface. For smaller flakes the embedding is complete, but for larger flakes a bending of the flake is induced. We rationalized this finding in terms of competing embedding and strain energies. While the edge C atoms interact strongly with the liquid in this embedded state, the central C atoms exhibit weak vdW-bonding to the surface at an adsorption height around or slightly above 3~$\text{\AA}$. Due to the weak interaction of the majority of the C atoms, the structural properties of the liquid surface (in terms of RDFs) are unperturbed by the presence of the flake.

We believe that the embedded structural motif is unique to liquid Cu graphene growth and speculate that it is key to understanding the special growth observed experimentally,
where both high growth speeds and low defect densities can be obtained simultaneously.\cite{Zheng2019,Saedi2020} A thorough analysis of this is a topic of our ongoing work. For example, we expect that barriers of the C precursor species to diffuse over the flake edge (Ehrlich-Schw\"{o}bel barrier) would be lowered in the embedded state. Also barriers for these species to attach to the flake edges may be significantly altered. The role of dissolved precursor species in combination with the embedded structural motif equivalently warrants further studies. Even if Cu has a very low C solubility compared to other common metal catalysts for graphene growth such as Ni, experiments have demonstrated that rather large quantities of C may be dissolved in the liquid Cu surface.\cite{Zeng2014} It is thereby puzzling that liquid Cu CVD produces single-layer graphene, since metals with a high C solubility are generally known to be susceptible to multilayer growth (in solid metal CVD).\cite{Bartelt2012} The embedded structural motif in liquid Cu CVD might be the key to resolving this puzzle.

\section{Supplementary material}
See the supplementary material for the correlation plot of force field and DFT forces, AIMD results for flake embedding and density profiles, the heights and charges of individual C atoms in the C$_{600}$ flake, and the 2D RDFs for Cu atoms located near the surface.

\begin{acknowledgments}
This project has received funding from the European Union's Horizon 2020 research and innovation programme under grant agreement 736299. Responsibility for the information and views set out in this article lies entirely with the authors. The authors gratefully acknowledge the Gauss Centre for Supercomputing e.V. (www.gauss-centre.eu) for funding this project by providing computing time through the John von Neumann Institute for Computing (NIC) on the GCS Supercomputer JUWELS \cite{JUWELS} at J{\"u}lich Supercomputing Centre (JSC).
\end{acknowledgments}

\section{Data availability}
The data that support the findings of this study are available from the corresponding author upon reasonable request.

\bibliography{MD_references}

\begin{thebibliography}{49}%
\makeatletter
\providecommand \@ifxundefined [1]{%
 \@ifx{#1\undefined}
}%
\providecommand \@ifnum [1]{%
 \ifnum #1\expandafter \@firstoftwo
 \else \expandafter \@secondoftwo
 \fi
}%
\providecommand \@ifx [1]{%
 \ifx #1\expandafter \@firstoftwo
 \else \expandafter \@secondoftwo
 \fi
}%
\providecommand \natexlab [1]{#1}%
\providecommand \enquote  [1]{``#1''}%
\providecommand \bibnamefont  [1]{#1}%
\providecommand \bibfnamefont [1]{#1}%
\providecommand \citenamefont [1]{#1}%
\providecommand \href@noop [0]{\@secondoftwo}%
\providecommand \href [0]{\begingroup \@sanitize@url \@href}%
\providecommand \@href[1]{\@@startlink{#1}\@@href}%
\providecommand \@@href[1]{\endgroup#1\@@endlink}%
\providecommand \@sanitize@url [0]{\catcode `\\12\catcode `\$12\catcode
  `\&12\catcode `\#12\catcode `\^12\catcode `\_12\catcode `\%12\relax}%
\providecommand \@@startlink[1]{}%
\providecommand \@@endlink[0]{}%
\providecommand \url  [0]{\begingroup\@sanitize@url \@url }%
\providecommand \@url [1]{\endgroup\@href {#1}{\urlprefix }}%
\providecommand \urlprefix  [0]{URL }%
\providecommand \Eprint [0]{\href }%
\providecommand \doibase [0]{http://dx.doi.org/}%
\providecommand \selectlanguage [0]{\@gobble}%
\providecommand \bibinfo  [0]{\@secondoftwo}%
\providecommand \bibfield  [0]{\@secondoftwo}%
\providecommand \translation [1]{[#1]}%
\providecommand \BibitemOpen [0]{}%
\providecommand \bibitemStop [0]{}%
\providecommand \bibitemNoStop [0]{.\EOS\space}%
\providecommand \EOS [0]{\spacefactor3000\relax}%
\providecommand \BibitemShut  [1]{\csname bibitem#1\endcsname}%
\let\auto@bib@innerbib\@empty
\bibitem [{\citenamefont {Geng}\ \emph {et~al.}(2012)\citenamefont {Geng},
  \citenamefont {Wu}, \citenamefont {Guo}, \citenamefont {Huang}, \citenamefont
  {Xue}, \citenamefont {Chen}, \citenamefont {Yu}, \citenamefont {Jiang},
  \citenamefont {Hu},\ and\ \citenamefont {Liu}}]{Geng2012}%
  \BibitemOpen
  \bibfield  {author} {\bibinfo {author} {\bibfnamefont {D.}~\bibnamefont
  {Geng}}, \bibinfo {author} {\bibfnamefont {B.}~\bibnamefont {Wu}}, \bibinfo
  {author} {\bibfnamefont {Y.}~\bibnamefont {Guo}}, \bibinfo {author}
  {\bibfnamefont {L.}~\bibnamefont {Huang}}, \bibinfo {author} {\bibfnamefont
  {Y.}~\bibnamefont {Xue}}, \bibinfo {author} {\bibfnamefont {J.}~\bibnamefont
  {Chen}}, \bibinfo {author} {\bibfnamefont {G.}~\bibnamefont {Yu}}, \bibinfo
  {author} {\bibfnamefont {L.}~\bibnamefont {Jiang}}, \bibinfo {author}
  {\bibfnamefont {W.}~\bibnamefont {Hu}}, \ and\ \bibinfo {author}
  {\bibfnamefont {Y.}~\bibnamefont {Liu}},\ }\href@noop {} {\bibfield
  {journal} {\bibinfo  {journal} {Proc. Natl. Acad. Sci. U.S.A.}\ }\textbf
  {\bibinfo {volume} {109}},\ \bibinfo {pages} {7992} (\bibinfo {year}
  {2012})}\BibitemShut {NoStop}%
\bibitem [{\citenamefont {Wu}\ \emph {et~al.}(2012)\citenamefont {Wu},
  \citenamefont {Fan}, \citenamefont {Speller}, \citenamefont {Creeth},
  \citenamefont {Sadowski}, \citenamefont {He}, \citenamefont {Robertson},
  \citenamefont {Allen},\ and\ \citenamefont {Warner}}]{Wu2012}%
  \BibitemOpen
  \bibfield  {author} {\bibinfo {author} {\bibfnamefont {Y.~A.}\ \bibnamefont
  {Wu}}, \bibinfo {author} {\bibfnamefont {Y.}~\bibnamefont {Fan}}, \bibinfo
  {author} {\bibfnamefont {S.}~\bibnamefont {Speller}}, \bibinfo {author}
  {\bibfnamefont {G.~L.}\ \bibnamefont {Creeth}}, \bibinfo {author}
  {\bibfnamefont {J.~T.}\ \bibnamefont {Sadowski}}, \bibinfo {author}
  {\bibfnamefont {K.}~\bibnamefont {He}}, \bibinfo {author} {\bibfnamefont
  {A.~W.}\ \bibnamefont {Robertson}}, \bibinfo {author} {\bibfnamefont {C.~S.}\
  \bibnamefont {Allen}}, \ and\ \bibinfo {author} {\bibfnamefont {J.~H.}\
  \bibnamefont {Warner}},\ }\href@noop {} {\bibfield  {journal} {\bibinfo
  {journal} {ACS Nano}\ }\textbf {\bibinfo {volume} {6}},\ \bibinfo {pages}
  {5010} (\bibinfo {year} {2012})}\BibitemShut {NoStop}%
\bibitem [{\citenamefont {Wu}\ \emph {et~al.}(2013)\citenamefont {Wu},
  \citenamefont {Geng}, \citenamefont {Xu}, \citenamefont {Guo}, \citenamefont
  {Huang}, \citenamefont {Xue}, \citenamefont {Chen}, \citenamefont {Yu},\ and\
  \citenamefont {Li}}]{Wu2013}%
  \BibitemOpen
  \bibfield  {author} {\bibinfo {author} {\bibfnamefont {B.}~\bibnamefont
  {Wu}}, \bibinfo {author} {\bibfnamefont {D.}~\bibnamefont {Geng}}, \bibinfo
  {author} {\bibfnamefont {Z.}~\bibnamefont {Xu}}, \bibinfo {author}
  {\bibfnamefont {Y.}~\bibnamefont {Guo}}, \bibinfo {author} {\bibfnamefont
  {L.}~\bibnamefont {Huang}}, \bibinfo {author} {\bibfnamefont
  {Y.}~\bibnamefont {Xue}}, \bibinfo {author} {\bibfnamefont {J.}~\bibnamefont
  {Chen}}, \bibinfo {author} {\bibfnamefont {G.}~\bibnamefont {Yu}}, \ and\
  \bibinfo {author} {\bibfnamefont {Z.}~\bibnamefont {Li}},\ }\href@noop {}
  {\bibfield  {journal} {\bibinfo  {journal} {NPG Asia Mater.}\ }\textbf
  {\bibinfo {volume} {5}},\ \bibinfo {pages} {e36} (\bibinfo {year}
  {2013})}\BibitemShut {NoStop}%
\bibitem [{\citenamefont {Geng}\ \emph {et~al.}(2014)\citenamefont {Geng},
  \citenamefont {Luo}, \citenamefont {Xu}, \citenamefont {Guo}, \citenamefont
  {Wu}, \citenamefont {Hu}, \citenamefont {Liu},\ and\ \citenamefont
  {Yu}}]{Geng2014}%
  \BibitemOpen
  \bibfield  {author} {\bibinfo {author} {\bibfnamefont {D.}~\bibnamefont
  {Geng}}, \bibinfo {author} {\bibfnamefont {B.}~\bibnamefont {Luo}}, \bibinfo
  {author} {\bibfnamefont {J.}~\bibnamefont {Xu}}, \bibinfo {author}
  {\bibfnamefont {Y.}~\bibnamefont {Guo}}, \bibinfo {author} {\bibfnamefont
  {B.}~\bibnamefont {Wu}}, \bibinfo {author} {\bibfnamefont {W.}~\bibnamefont
  {Hu}}, \bibinfo {author} {\bibfnamefont {Y.}~\bibnamefont {Liu}}, \ and\
  \bibinfo {author} {\bibfnamefont {G.}~\bibnamefont {Yu}},\ }\href@noop {}
  {\bibfield  {journal} {\bibinfo  {journal} {Adv. Funct. Mater.}\ }\textbf
  {\bibinfo {volume} {24}},\ \bibinfo {pages} {1664} (\bibinfo {year}
  {2014})}\BibitemShut {NoStop}%
\bibitem [{\citenamefont {Zeng}\ \emph {et~al.}(2016)\citenamefont {Zeng},
  \citenamefont {Wang}, \citenamefont {Liu}, \citenamefont {Zhang},
  \citenamefont {Xue}, \citenamefont {Xiao}, \citenamefont {Qin},\ and\
  \citenamefont {Fu}}]{Zeng2016}%
  \BibitemOpen
  \bibfield  {author} {\bibinfo {author} {\bibfnamefont {M.}~\bibnamefont
  {Zeng}}, \bibinfo {author} {\bibfnamefont {L.}~\bibnamefont {Wang}}, \bibinfo
  {author} {\bibfnamefont {J.}~\bibnamefont {Liu}}, \bibinfo {author}
  {\bibfnamefont {T.}~\bibnamefont {Zhang}}, \bibinfo {author} {\bibfnamefont
  {H.}~\bibnamefont {Xue}}, \bibinfo {author} {\bibfnamefont {Y.}~\bibnamefont
  {Xiao}}, \bibinfo {author} {\bibfnamefont {Z.}~\bibnamefont {Qin}}, \ and\
  \bibinfo {author} {\bibfnamefont {L.}~\bibnamefont {Fu}},\ }\href@noop {}
  {\bibfield  {journal} {\bibinfo  {journal} {J. Am. Chem. Soc.}\ }\textbf
  {\bibinfo {volume} {138}},\ \bibinfo {pages} {7812} (\bibinfo {year}
  {2016})}\BibitemShut {NoStop}%
\bibitem [{\citenamefont {Xue}\ \emph {et~al.}(2019)\citenamefont {Xue},
  \citenamefont {Xu}, \citenamefont {Wang}, \citenamefont {Liu}, \citenamefont
  {Jiang}, \citenamefont {Yu}, \citenamefont {Zhou}, \citenamefont {Ma},
  \citenamefont {Wang},\ and\ \citenamefont {Yu}}]{Xue2019}%
  \BibitemOpen
  \bibfield  {author} {\bibinfo {author} {\bibfnamefont {X.}~\bibnamefont
  {Xue}}, \bibinfo {author} {\bibfnamefont {Q.}~\bibnamefont {Xu}}, \bibinfo
  {author} {\bibfnamefont {H.}~\bibnamefont {Wang}}, \bibinfo {author}
  {\bibfnamefont {S.}~\bibnamefont {Liu}}, \bibinfo {author} {\bibfnamefont
  {Q.}~\bibnamefont {Jiang}}, \bibinfo {author} {\bibfnamefont
  {Z.}~\bibnamefont {Yu}}, \bibinfo {author} {\bibfnamefont {X.}~\bibnamefont
  {Zhou}}, \bibinfo {author} {\bibfnamefont {T.}~\bibnamefont {Ma}}, \bibinfo
  {author} {\bibfnamefont {L.}~\bibnamefont {Wang}}, \ and\ \bibinfo {author}
  {\bibfnamefont {G.}~\bibnamefont {Yu}},\ }\href@noop {} {\bibfield  {journal}
  {\bibinfo  {journal} {Chem. Mater.}\ }\textbf {\bibinfo {volume} {31}},\
  \bibinfo {pages} {1231} (\bibinfo {year} {2019})}\BibitemShut {NoStop}%
\bibitem [{\citenamefont {Zheng}\ \emph {et~al.}(2019)\citenamefont {Zheng},
  \citenamefont {Zeng}, \citenamefont {Cao}, \citenamefont {Zhang},
  \citenamefont {Gao}, \citenamefont {Xiao},\ and\ \citenamefont
  {Fu}}]{Zheng2019}%
  \BibitemOpen
  \bibfield  {author} {\bibinfo {author} {\bibfnamefont {S.}~\bibnamefont
  {Zheng}}, \bibinfo {author} {\bibfnamefont {M.}~\bibnamefont {Zeng}},
  \bibinfo {author} {\bibfnamefont {H.}~\bibnamefont {Cao}}, \bibinfo {author}
  {\bibfnamefont {T.}~\bibnamefont {Zhang}}, \bibinfo {author} {\bibfnamefont
  {X.}~\bibnamefont {Gao}}, \bibinfo {author} {\bibfnamefont {Y.}~\bibnamefont
  {Xiao}}, \ and\ \bibinfo {author} {\bibfnamefont {L.}~\bibnamefont {Fu}},\
  }\href@noop {} {\bibfield  {journal} {\bibinfo  {journal} {Science China
  Mat.}\ }\textbf {\bibinfo {volume} {62}},\ \bibinfo {pages} {1087} (\bibinfo
  {year} {2019})}\BibitemShut {NoStop}%
\bibitem [{\citenamefont {Xu}\ \emph {et~al.}(2017)\citenamefont {Xu},
  \citenamefont {Zhang}, \citenamefont {Dong}, \citenamefont {Yi},
  \citenamefont {Niu}, \citenamefont {Wu}, \citenamefont {Lin}, \citenamefont
  {Yin}, \citenamefont {Li}, \citenamefont {Zhou}, \citenamefont {Wang},
  \citenamefont {Sun}, \citenamefont {Duan}, \citenamefont {Gao}, \citenamefont
  {Jiang}, \citenamefont {Wu}, \citenamefont {Peng}, \citenamefont {Ruoff},
  \citenamefont {Liu}, \citenamefont {Yu}, \citenamefont {Wang}, \citenamefont
  {Ding},\ and\ \citenamefont {Liu}}]{Xu2017}%
  \BibitemOpen
  \bibfield  {author} {\bibinfo {author} {\bibfnamefont {X.}~\bibnamefont
  {Xu}}, \bibinfo {author} {\bibfnamefont {Z.}~\bibnamefont {Zhang}}, \bibinfo
  {author} {\bibfnamefont {J.}~\bibnamefont {Dong}}, \bibinfo {author}
  {\bibfnamefont {D.}~\bibnamefont {Yi}}, \bibinfo {author} {\bibfnamefont
  {J.}~\bibnamefont {Niu}}, \bibinfo {author} {\bibfnamefont {M.}~\bibnamefont
  {Wu}}, \bibinfo {author} {\bibfnamefont {L.}~\bibnamefont {Lin}}, \bibinfo
  {author} {\bibfnamefont {R.}~\bibnamefont {Yin}}, \bibinfo {author}
  {\bibfnamefont {M.}~\bibnamefont {Li}}, \bibinfo {author} {\bibfnamefont
  {J.}~\bibnamefont {Zhou}}, \bibinfo {author} {\bibfnamefont {S.}~\bibnamefont
  {Wang}}, \bibinfo {author} {\bibfnamefont {J.}~\bibnamefont {Sun}}, \bibinfo
  {author} {\bibfnamefont {X.}~\bibnamefont {Duan}}, \bibinfo {author}
  {\bibfnamefont {P.}~\bibnamefont {Gao}}, \bibinfo {author} {\bibfnamefont
  {Y.}~\bibnamefont {Jiang}}, \bibinfo {author} {\bibfnamefont
  {X.}~\bibnamefont {Wu}}, \bibinfo {author} {\bibfnamefont {H.}~\bibnamefont
  {Peng}}, \bibinfo {author} {\bibfnamefont {R.~S.}\ \bibnamefont {Ruoff}},
  \bibinfo {author} {\bibfnamefont {Z.}~\bibnamefont {Liu}}, \bibinfo {author}
  {\bibfnamefont {D.}~\bibnamefont {Yu}}, \bibinfo {author} {\bibfnamefont
  {E.}~\bibnamefont {Wang}}, \bibinfo {author} {\bibfnamefont {F.}~\bibnamefont
  {Ding}}, \ and\ \bibinfo {author} {\bibfnamefont {K.}~\bibnamefont {Liu}},\
  }\href@noop {} {\bibfield  {journal} {\bibinfo  {journal} {Sci. Bull.}\
  }\textbf {\bibinfo {volume} {62}},\ \bibinfo {pages} {1074 } (\bibinfo {year}
  {2017})}\BibitemShut {NoStop}%
\bibitem [{\citenamefont {Nguyen}\ \emph {et~al.}(2015)\citenamefont {Nguyen},
  \citenamefont {Shin}, \citenamefont {Duong}, \citenamefont {Kim},
  \citenamefont {Perello}, \citenamefont {Lim}, \citenamefont {Yuan},
  \citenamefont {Ding}, \citenamefont {Jeong}, \citenamefont {Shin},
  \citenamefont {Lee}, \citenamefont {Chae}, \citenamefont {Vu}, \citenamefont
  {Lee},\ and\ \citenamefont {Lee}}]{Nguyen2015}%
  \BibitemOpen
  \bibfield  {author} {\bibinfo {author} {\bibfnamefont {V.~L.}\ \bibnamefont
  {Nguyen}}, \bibinfo {author} {\bibfnamefont {B.~G.}\ \bibnamefont {Shin}},
  \bibinfo {author} {\bibfnamefont {D.~L.}\ \bibnamefont {Duong}}, \bibinfo
  {author} {\bibfnamefont {S.~T.}\ \bibnamefont {Kim}}, \bibinfo {author}
  {\bibfnamefont {D.}~\bibnamefont {Perello}}, \bibinfo {author} {\bibfnamefont
  {Y.~J.}\ \bibnamefont {Lim}}, \bibinfo {author} {\bibfnamefont {Q.~H.}\
  \bibnamefont {Yuan}}, \bibinfo {author} {\bibfnamefont {F.}~\bibnamefont
  {Ding}}, \bibinfo {author} {\bibfnamefont {H.~Y.}\ \bibnamefont {Jeong}},
  \bibinfo {author} {\bibfnamefont {H.~S.}\ \bibnamefont {Shin}}, \bibinfo
  {author} {\bibfnamefont {S.~M.}\ \bibnamefont {Lee}}, \bibinfo {author}
  {\bibfnamefont {S.~H.}\ \bibnamefont {Chae}}, \bibinfo {author}
  {\bibfnamefont {Q.~A.}\ \bibnamefont {Vu}}, \bibinfo {author} {\bibfnamefont
  {S.~H.}\ \bibnamefont {Lee}}, \ and\ \bibinfo {author} {\bibfnamefont
  {Y.~H.}\ \bibnamefont {Lee}},\ }\href@noop {} {\bibfield  {journal} {\bibinfo
   {journal} {Adv. Mater.}\ }\textbf {\bibinfo {volume} {27}},\ \bibinfo
  {pages} {1376} (\bibinfo {year} {2015})}\BibitemShut {NoStop}%
\bibitem [{\citenamefont {Gao}, \citenamefont {Guest},\ and\ \citenamefont
  {Guisinger}(2010)}]{Gao2010}%
  \BibitemOpen
  \bibfield  {author} {\bibinfo {author} {\bibfnamefont {L.}~\bibnamefont
  {Gao}}, \bibinfo {author} {\bibfnamefont {J.~R.}\ \bibnamefont {Guest}}, \
  and\ \bibinfo {author} {\bibfnamefont {N.~P.}\ \bibnamefont {Guisinger}},\
  }\href@noop {} {\bibfield  {journal} {\bibinfo  {journal} {Nano Lett.}\
  }\textbf {\bibinfo {volume} {10}},\ \bibinfo {pages} {3512} (\bibinfo {year}
  {2010})}\BibitemShut {NoStop}%
\bibitem [{\citenamefont {Li}\ \emph {et~al.}(2009)\citenamefont {Li},
  \citenamefont {Cai}, \citenamefont {An}, \citenamefont {Kim}, \citenamefont
  {Nah}, \citenamefont {Yang}, \citenamefont {Piner}, \citenamefont
  {Velamakanni}, \citenamefont {Jung}, \citenamefont {Tutuc}, \citenamefont
  {Banerjee}, \citenamefont {Colombo},\ and\ \citenamefont {Ruoff}}]{Li2009}%
  \BibitemOpen
  \bibfield  {author} {\bibinfo {author} {\bibfnamefont {X.}~\bibnamefont
  {Li}}, \bibinfo {author} {\bibfnamefont {W.}~\bibnamefont {Cai}}, \bibinfo
  {author} {\bibfnamefont {J.}~\bibnamefont {An}}, \bibinfo {author}
  {\bibfnamefont {S.}~\bibnamefont {Kim}}, \bibinfo {author} {\bibfnamefont
  {J.}~\bibnamefont {Nah}}, \bibinfo {author} {\bibfnamefont {D.}~\bibnamefont
  {Yang}}, \bibinfo {author} {\bibfnamefont {R.}~\bibnamefont {Piner}},
  \bibinfo {author} {\bibfnamefont {A.}~\bibnamefont {Velamakanni}}, \bibinfo
  {author} {\bibfnamefont {I.}~\bibnamefont {Jung}}, \bibinfo {author}
  {\bibfnamefont {E.}~\bibnamefont {Tutuc}}, \bibinfo {author} {\bibfnamefont
  {S.~K.}\ \bibnamefont {Banerjee}}, \bibinfo {author} {\bibfnamefont
  {L.}~\bibnamefont {Colombo}}, \ and\ \bibinfo {author} {\bibfnamefont
  {R.~S.}\ \bibnamefont {Ruoff}},\ }\href@noop {} {\bibfield  {journal}
  {\bibinfo  {journal} {Science}\ }\textbf {\bibinfo {volume} {324}},\ \bibinfo
  {pages} {1312} (\bibinfo {year} {2009})}\BibitemShut {NoStop}%
\bibitem [{\citenamefont {Bhaviripudi}\ \emph {et~al.}(2010)\citenamefont
  {Bhaviripudi}, \citenamefont {Jia}, \citenamefont {Dresselhaus},\ and\
  \citenamefont {Kong}}]{Bhaviripudi2010}%
  \BibitemOpen
  \bibfield  {author} {\bibinfo {author} {\bibfnamefont {S.}~\bibnamefont
  {Bhaviripudi}}, \bibinfo {author} {\bibfnamefont {X.}~\bibnamefont {Jia}},
  \bibinfo {author} {\bibfnamefont {M.~S.}\ \bibnamefont {Dresselhaus}}, \ and\
  \bibinfo {author} {\bibfnamefont {J.}~\bibnamefont {Kong}},\ }\href@noop {}
  {\bibfield  {journal} {\bibinfo  {journal} {Nano Lett.}\ }\textbf {\bibinfo
  {volume} {10}},\ \bibinfo {pages} {4128} (\bibinfo {year}
  {2010})}\BibitemShut {NoStop}%
\bibitem [{\citenamefont {Vlassiouk}\ \emph {et~al.}(2011)\citenamefont
  {Vlassiouk}, \citenamefont {Regmi}, \citenamefont {Fulvio}, \citenamefont
  {Dai}, \citenamefont {Datskos}, \citenamefont {Eres},\ and\ \citenamefont
  {Smirnov}}]{Vlassiouk2011}%
  \BibitemOpen
  \bibfield  {author} {\bibinfo {author} {\bibfnamefont {I.}~\bibnamefont
  {Vlassiouk}}, \bibinfo {author} {\bibfnamefont {M.}~\bibnamefont {Regmi}},
  \bibinfo {author} {\bibfnamefont {P.}~\bibnamefont {Fulvio}}, \bibinfo
  {author} {\bibfnamefont {S.}~\bibnamefont {Dai}}, \bibinfo {author}
  {\bibfnamefont {P.}~\bibnamefont {Datskos}}, \bibinfo {author} {\bibfnamefont
  {G.}~\bibnamefont {Eres}}, \ and\ \bibinfo {author} {\bibfnamefont
  {S.}~\bibnamefont {Smirnov}},\ }\href@noop {} {\bibfield  {journal} {\bibinfo
   {journal} {ACS Nano}\ }\textbf {\bibinfo {volume} {5}},\ \bibinfo {pages}
  {6069} (\bibinfo {year} {2011})}\BibitemShut {NoStop}%
\bibitem [{\citenamefont {Kim}\ \emph {et~al.}(2012)\citenamefont {Kim},
  \citenamefont {Mattevi}, \citenamefont {Calvo}, \citenamefont {Oberg},
  \citenamefont {Artiglia}, \citenamefont {Agnoli}, \citenamefont
  {Hirjibehedin}, \citenamefont {Chhowalla},\ and\ \citenamefont
  {Saiz}}]{Kim2012}%
  \BibitemOpen
  \bibfield  {author} {\bibinfo {author} {\bibfnamefont {H.}~\bibnamefont
  {Kim}}, \bibinfo {author} {\bibfnamefont {C.}~\bibnamefont {Mattevi}},
  \bibinfo {author} {\bibfnamefont {M.~R.}\ \bibnamefont {Calvo}}, \bibinfo
  {author} {\bibfnamefont {J.~C.}\ \bibnamefont {Oberg}}, \bibinfo {author}
  {\bibfnamefont {L.}~\bibnamefont {Artiglia}}, \bibinfo {author}
  {\bibfnamefont {S.}~\bibnamefont {Agnoli}}, \bibinfo {author} {\bibfnamefont
  {C.~F.}\ \bibnamefont {Hirjibehedin}}, \bibinfo {author} {\bibfnamefont
  {M.}~\bibnamefont {Chhowalla}}, \ and\ \bibinfo {author} {\bibfnamefont
  {E.}~\bibnamefont {Saiz}},\ }\href@noop {} {\bibfield  {journal} {\bibinfo
  {journal} {ACS Nano}\ }\textbf {\bibinfo {volume} {6}},\ \bibinfo {pages}
  {3614} (\bibinfo {year} {2012})}\BibitemShut {NoStop}%
\bibitem [{\citenamefont {Li}, \citenamefont {Colombo},\ and\ \citenamefont
  {Ruoff}(2016)}]{Li2016}%
  \BibitemOpen
  \bibfield  {author} {\bibinfo {author} {\bibfnamefont {X.}~\bibnamefont
  {Li}}, \bibinfo {author} {\bibfnamefont {L.}~\bibnamefont {Colombo}}, \ and\
  \bibinfo {author} {\bibfnamefont {R.~S.}\ \bibnamefont {Ruoff}},\ }\href@noop
  {} {\bibfield  {journal} {\bibinfo  {journal} {Adv. Mater.}\ }\textbf
  {\bibinfo {volume} {28}},\ \bibinfo {pages} {6247} (\bibinfo {year}
  {2016})}\BibitemShut {NoStop}%
\bibitem [{\citenamefont {Huet}\ and\ \citenamefont {Raskin}(2017)}]{Huet2017}%
  \BibitemOpen
  \bibfield  {author} {\bibinfo {author} {\bibfnamefont {B.}~\bibnamefont
  {Huet}}\ and\ \bibinfo {author} {\bibfnamefont {J.-P.}\ \bibnamefont
  {Raskin}},\ }\href@noop {} {\bibfield  {journal} {\bibinfo  {journal} {Chem.
  Mater.}\ }\textbf {\bibinfo {volume} {29}},\ \bibinfo {pages} {3431}
  (\bibinfo {year} {2017})}\BibitemShut {NoStop}%
\bibitem [{\citenamefont {Novoselov}\ \emph {et~al.}(2005)\citenamefont
  {Novoselov}, \citenamefont {Geim}, \citenamefont {Morozov}, \citenamefont
  {Jiang}, \citenamefont {Katsnelson}, \citenamefont {Grigorieva},
  \citenamefont {Dubonos},\ and\ \citenamefont {Firsov}}]{Novoselov2005}%
  \BibitemOpen
  \bibfield  {author} {\bibinfo {author} {\bibfnamefont {K.}~\bibnamefont
  {Novoselov}}, \bibinfo {author} {\bibfnamefont {A.}~\bibnamefont {Geim}},
  \bibinfo {author} {\bibfnamefont {S.}~\bibnamefont {Morozov}}, \bibinfo
  {author} {\bibfnamefont {D.}~\bibnamefont {Jiang}}, \bibinfo {author}
  {\bibfnamefont {M.}~\bibnamefont {Katsnelson}}, \bibinfo {author}
  {\bibfnamefont {I.}~\bibnamefont {Grigorieva}}, \bibinfo {author}
  {\bibfnamefont {S.}~\bibnamefont {Dubonos}}, \ and\ \bibinfo {author}
  {\bibfnamefont {A.}~\bibnamefont {Firsov}},\ }\href@noop {} {\bibfield
  {journal} {\bibinfo  {journal} {Nature}\ }\textbf {\bibinfo {volume} {438}},\
  \bibinfo {pages} {197} (\bibinfo {year} {2005})}\BibitemShut {NoStop}%
\bibitem [{\citenamefont {Allen}, \citenamefont {Tung},\ and\ \citenamefont
  {Kaner}(2010)}]{Allen2010}%
  \BibitemOpen
  \bibfield  {author} {\bibinfo {author} {\bibfnamefont {M.~J.}\ \bibnamefont
  {Allen}}, \bibinfo {author} {\bibfnamefont {V.~C.}\ \bibnamefont {Tung}}, \
  and\ \bibinfo {author} {\bibfnamefont {R.~B.}\ \bibnamefont {Kaner}},\
  }\href@noop {} {\bibfield  {journal} {\bibinfo  {journal} {Chem. Rev.}\
  }\textbf {\bibinfo {volume} {110}},\ \bibinfo {pages} {132} (\bibinfo {year}
  {2010})}\BibitemShut {NoStop}%
\bibitem [{\citenamefont {Saedi}\ \emph {et~al.}(2020)\citenamefont {Saedi},
  \citenamefont {de~Voogd}, \citenamefont {Sjardin}, \citenamefont {Manikas},
  \citenamefont {Galiotis}, \citenamefont {Jankowski}, \citenamefont {Renaud},
  \citenamefont {La~Porta}, \citenamefont {Konovalov}, \citenamefont {van
  Baarle},\ and\ \citenamefont {Groot}}]{Saedi2020}%
  \BibitemOpen
  \bibfield  {author} {\bibinfo {author} {\bibfnamefont {M.}~\bibnamefont
  {Saedi}}, \bibinfo {author} {\bibfnamefont {J.~M.}\ \bibnamefont {de~Voogd}},
  \bibinfo {author} {\bibfnamefont {A.}~\bibnamefont {Sjardin}}, \bibinfo
  {author} {\bibfnamefont {A.}~\bibnamefont {Manikas}}, \bibinfo {author}
  {\bibfnamefont {C.}~\bibnamefont {Galiotis}}, \bibinfo {author}
  {\bibfnamefont {M.}~\bibnamefont {Jankowski}}, \bibinfo {author}
  {\bibfnamefont {G.}~\bibnamefont {Renaud}}, \bibinfo {author} {\bibfnamefont
  {F.}~\bibnamefont {La~Porta}}, \bibinfo {author} {\bibfnamefont
  {O.}~\bibnamefont {Konovalov}}, \bibinfo {author} {\bibfnamefont {G.~J.~C.}\
  \bibnamefont {van Baarle}}, \ and\ \bibinfo {author} {\bibfnamefont
  {I.~M.~N.}\ \bibnamefont {Groot}},\ }\href@noop {} {\bibfield  {journal}
  {\bibinfo  {journal} {Rev. Sci. Instrum.}\ }\textbf {\bibinfo {volume}
  {91}},\ \bibinfo {pages} {013907} (\bibinfo {year} {2020})}\BibitemShut
  {NoStop}%
\bibitem [{\citenamefont {Jankowski}\ \emph {et~al.}()\citenamefont
  {Jankowski}, \citenamefont {Porta}, \citenamefont {Manikas}, \citenamefont
  {Tsakonas}, \citenamefont {Cingolani}, \citenamefont {Andersen},
  \citenamefont {de~Voogd}, \citenamefont {van Baarle}, \citenamefont {Reuter},
  \citenamefont {Galiotis}, \citenamefont {Groot}, \citenamefont {Renaud},
  \citenamefont {Konovalov},\ and\ \citenamefont {Saedi}}]{SciencePaper}%
  \BibitemOpen
  \bibfield  {author} {\bibinfo {author} {\bibfnamefont {M.}~\bibnamefont
  {Jankowski}}, \bibinfo {author} {\bibfnamefont {F.~L.}\ \bibnamefont
  {Porta}}, \bibinfo {author} {\bibfnamefont {A.}~\bibnamefont {Manikas}},
  \bibinfo {author} {\bibfnamefont {C.}~\bibnamefont {Tsakonas}}, \bibinfo
  {author} {\bibfnamefont {J.~S.}\ \bibnamefont {Cingolani}}, \bibinfo {author}
  {\bibfnamefont {M.}~\bibnamefont {Andersen}}, \bibinfo {author}
  {\bibfnamefont {J.~M.}\ \bibnamefont {de~Voogd}}, \bibinfo {author}
  {\bibfnamefont {G.~J.~C.}\ \bibnamefont {van Baarle}}, \bibinfo {author}
  {\bibfnamefont {K.}~\bibnamefont {Reuter}}, \bibinfo {author} {\bibfnamefont
  {C.}~\bibnamefont {Galiotis}}, \bibinfo {author} {\bibfnamefont {I.~M.~N.}\
  \bibnamefont {Groot}}, \bibinfo {author} {\bibfnamefont {G.}~\bibnamefont
  {Renaud}}, \bibinfo {author} {\bibfnamefont {O.}~\bibnamefont {Konovalov}}, \
  and\ \bibinfo {author} {\bibfnamefont {A.}~\bibnamefont {Saedi}},\
  }\href@noop {} {\bibinfo  {journal} {submitted}\ }\BibitemShut {NoStop}%
\bibitem [{\citenamefont {Li}\ \emph {et~al.}(2014)\citenamefont {Li},
  \citenamefont {Page}, \citenamefont {Hettich}, \citenamefont {Aradi},
  \citenamefont {Kohler}, \citenamefont {Frauenheim}, \citenamefont {Irle},\
  and\ \citenamefont {Morokuma}}]{Li2014}%
  \BibitemOpen
\bibfield  {journal} {  }\bibfield  {author} {\bibinfo {author} {\bibfnamefont
  {H.-B.}\ \bibnamefont {Li}}, \bibinfo {author} {\bibfnamefont {A.~J.}\
  \bibnamefont {Page}}, \bibinfo {author} {\bibfnamefont {C.}~\bibnamefont
  {Hettich}}, \bibinfo {author} {\bibfnamefont {B.}~\bibnamefont {Aradi}},
  \bibinfo {author} {\bibfnamefont {C.}~\bibnamefont {Kohler}}, \bibinfo
  {author} {\bibfnamefont {T.}~\bibnamefont {Frauenheim}}, \bibinfo {author}
  {\bibfnamefont {S.}~\bibnamefont {Irle}}, \ and\ \bibinfo {author}
  {\bibfnamefont {K.}~\bibnamefont {Morokuma}},\ }\href@noop {} {\bibfield
  {journal} {\bibinfo  {journal} {Chem. Sci.}\ }\textbf {\bibinfo {volume}
  {5}},\ \bibinfo {pages} {3493} (\bibinfo {year} {2014})}\BibitemShut
  {NoStop}%
\bibitem [{\citenamefont {Liang}\ \emph {et~al.}(2012)\citenamefont {Liang},
  \citenamefont {Devine}, \citenamefont {Phillpot},\ and\ \citenamefont
  {Sinnott}}]{Liang2012}%
  \BibitemOpen
  \bibfield  {author} {\bibinfo {author} {\bibfnamefont {T.}~\bibnamefont
  {Liang}}, \bibinfo {author} {\bibfnamefont {B.}~\bibnamefont {Devine}},
  \bibinfo {author} {\bibfnamefont {S.~R.}\ \bibnamefont {Phillpot}}, \ and\
  \bibinfo {author} {\bibfnamefont {S.~B.}\ \bibnamefont {Sinnott}},\
  }\href@noop {} {\bibfield  {journal} {\bibinfo  {journal} {J. Phys. Chem. A}\
  }\textbf {\bibinfo {volume} {116}},\ \bibinfo {pages} {7976} (\bibinfo {year}
  {2012})}\BibitemShut {NoStop}%
\bibitem [{\citenamefont {Liang}\ \emph {et~al.}(2013)\citenamefont {Liang},
  \citenamefont {Shan}, \citenamefont {Cheng}, \citenamefont {Devine},
  \citenamefont {Noordhoek}, \citenamefont {Li}, \citenamefont {Lu},
  \citenamefont {Phillpot},\ and\ \citenamefont {Sinnott}}]{Liang2013}%
  \BibitemOpen
  \bibfield  {author} {\bibinfo {author} {\bibfnamefont {T.}~\bibnamefont
  {Liang}}, \bibinfo {author} {\bibfnamefont {T.-R.}\ \bibnamefont {Shan}},
  \bibinfo {author} {\bibfnamefont {Y.-T.}\ \bibnamefont {Cheng}}, \bibinfo
  {author} {\bibfnamefont {B.~D.}\ \bibnamefont {Devine}}, \bibinfo {author}
  {\bibfnamefont {M.}~\bibnamefont {Noordhoek}}, \bibinfo {author}
  {\bibfnamefont {Y.}~\bibnamefont {Li}}, \bibinfo {author} {\bibfnamefont
  {Z.}~\bibnamefont {Lu}}, \bibinfo {author} {\bibfnamefont {S.~R.}\
  \bibnamefont {Phillpot}}, \ and\ \bibinfo {author} {\bibfnamefont {S.~B.}\
  \bibnamefont {Sinnott}},\ }\href@noop {} {\bibfield  {journal} {\bibinfo
  {journal} {Mater. Sci. Eng. R Rep.}\ }\textbf {\bibinfo {volume} {74}},\
  \bibinfo {pages} {255 } (\bibinfo {year} {2013})}\BibitemShut {NoStop}%
\bibitem [{\citenamefont {Klaver}\ \emph {et~al.}(2015)\citenamefont {Klaver},
  \citenamefont {Zhu}, \citenamefont {Sluiter},\ and\ \citenamefont
  {Janssen}}]{Klaver2015}%
  \BibitemOpen
  \bibfield  {author} {\bibinfo {author} {\bibfnamefont {T.}~\bibnamefont
  {Klaver}}, \bibinfo {author} {\bibfnamefont {S.-E.}\ \bibnamefont {Zhu}},
  \bibinfo {author} {\bibfnamefont {M.}~\bibnamefont {Sluiter}}, \ and\
  \bibinfo {author} {\bibfnamefont {G.}~\bibnamefont {Janssen}},\ }\href@noop
  {} {\bibfield  {journal} {\bibinfo  {journal} {Carbon}\ }\textbf {\bibinfo
  {volume} {82}},\ \bibinfo {pages} {538 } (\bibinfo {year}
  {2015})}\BibitemShut {NoStop}%
\bibitem [{\citenamefont {Xu}\ \emph {et~al.}(2020)\citenamefont {Xu},
  \citenamefont {Zhao}, \citenamefont {Qiu}, \citenamefont {Zhang},
  \citenamefont {Qiao},\ and\ \citenamefont {Ding}}]{Xu2020}%
  \BibitemOpen
  \bibfield  {author} {\bibinfo {author} {\bibfnamefont {Z.}~\bibnamefont
  {Xu}}, \bibinfo {author} {\bibfnamefont {G.}~\bibnamefont {Zhao}}, \bibinfo
  {author} {\bibfnamefont {L.}~\bibnamefont {Qiu}}, \bibinfo {author}
  {\bibfnamefont {X.}~\bibnamefont {Zhang}}, \bibinfo {author} {\bibfnamefont
  {G.}~\bibnamefont {Qiao}}, \ and\ \bibinfo {author} {\bibfnamefont
  {F.}~\bibnamefont {Ding}},\ }\href@noop {} {\bibfield  {journal} {\bibinfo
  {journal} {Npj Comput. Mater.}\ }\textbf {\bibinfo {volume} {6}},\ \bibinfo
  {pages} {14} (\bibinfo {year} {2020})}\BibitemShut {NoStop}%
\bibitem [{\citenamefont {Yuan}, \citenamefont {Yakobson},\ and\ \citenamefont
  {Ding}(2014)}]{Yuan2014}%
  \BibitemOpen
  \bibfield  {author} {\bibinfo {author} {\bibfnamefont {Q.}~\bibnamefont
  {Yuan}}, \bibinfo {author} {\bibfnamefont {B.~I.}\ \bibnamefont {Yakobson}},
  \ and\ \bibinfo {author} {\bibfnamefont {F.}~\bibnamefont {Ding}},\
  }\href@noop {} {\bibfield  {journal} {\bibinfo  {journal} {J. Phys. Chem.
  Lett.}\ }\textbf {\bibinfo {volume} {5}},\ \bibinfo {pages} {3093} (\bibinfo
  {year} {2014})}\BibitemShut {NoStop}%
\bibitem [{\citenamefont {Pozzo}\ \emph {et~al.}(2011)\citenamefont {Pozzo},
  \citenamefont {Alf\`e}, \citenamefont {Lacovig}, \citenamefont {Hofmann},
  \citenamefont {Lizzit},\ and\ \citenamefont {Baraldi}}]{Pozzo2011}%
  \BibitemOpen
  \bibfield  {author} {\bibinfo {author} {\bibfnamefont {M.}~\bibnamefont
  {Pozzo}}, \bibinfo {author} {\bibfnamefont {D.}~\bibnamefont {Alf\`e}},
  \bibinfo {author} {\bibfnamefont {P.}~\bibnamefont {Lacovig}}, \bibinfo
  {author} {\bibfnamefont {P.}~\bibnamefont {Hofmann}}, \bibinfo {author}
  {\bibfnamefont {S.}~\bibnamefont {Lizzit}}, \ and\ \bibinfo {author}
  {\bibfnamefont {A.}~\bibnamefont {Baraldi}},\ }\href@noop {} {\bibfield
  {journal} {\bibinfo  {journal} {Phys. Rev. Lett.}\ }\textbf {\bibinfo
  {volume} {106}},\ \bibinfo {pages} {135501} (\bibinfo {year}
  {2011})}\BibitemShut {NoStop}%
\bibitem [{\citenamefont {Plimpton}(1995)}]{Plimpton1995}%
  \BibitemOpen
  \bibfield  {author} {\bibinfo {author} {\bibfnamefont {S.}~\bibnamefont
  {Plimpton}},\ }\href@noop {} {\bibfield  {journal} {\bibinfo  {journal} {J.
  Comput. Phys.}\ }\textbf {\bibinfo {volume} {117}},\ \bibinfo {pages} {1 }
  (\bibinfo {year} {1995})}\BibitemShut {NoStop}%
\bibitem [{\citenamefont {Larsen}\ \emph {et~al.}(2017)\citenamefont {Larsen},
  \citenamefont {Mortensen}, \citenamefont {Blomqvist}, \citenamefont
  {Castelli}, \citenamefont {Christensen}, \citenamefont {Du{\l}ak},
  \citenamefont {Friis}, \citenamefont {Groves}, \citenamefont {Hammer},
  \citenamefont {Hargus}, \citenamefont {Hermes}, \citenamefont {Jennings},
  \citenamefont {Jensen}, \citenamefont {Kermode}, \citenamefont {Kitchin},
  \citenamefont {Kolsbjerg}, \citenamefont {Kubal}, \citenamefont {Kaasbjerg},
  \citenamefont {Lysgaard}, \citenamefont {Maronsson}, \citenamefont {Maxson},
  \citenamefont {Olsen}, \citenamefont {Pastewka}, \citenamefont {Peterson},
  \citenamefont {Rostgaard}, \citenamefont {Schi{\o}tz}, \citenamefont
  {Schütt}, \citenamefont {Strange}, \citenamefont {Thygesen}, \citenamefont
  {Vegge}, \citenamefont {Vilhelmsen}, \citenamefont {Walter}, \citenamefont
  {Zeng},\ and\ \citenamefont {Jacobsen}}]{Hjorth_Larsen_2017}%
  \BibitemOpen
  \bibfield  {author} {\bibinfo {author} {\bibfnamefont {A.~H.}\ \bibnamefont
  {Larsen}}, \bibinfo {author} {\bibfnamefont {J.~J.}\ \bibnamefont
  {Mortensen}}, \bibinfo {author} {\bibfnamefont {J.}~\bibnamefont
  {Blomqvist}}, \bibinfo {author} {\bibfnamefont {I.~E.}\ \bibnamefont
  {Castelli}}, \bibinfo {author} {\bibfnamefont {R.}~\bibnamefont
  {Christensen}}, \bibinfo {author} {\bibfnamefont {M.}~\bibnamefont
  {Du{\l}ak}}, \bibinfo {author} {\bibfnamefont {J.}~\bibnamefont {Friis}},
  \bibinfo {author} {\bibfnamefont {M.~N.}\ \bibnamefont {Groves}}, \bibinfo
  {author} {\bibfnamefont {B.}~\bibnamefont {Hammer}}, \bibinfo {author}
  {\bibfnamefont {C.}~\bibnamefont {Hargus}}, \bibinfo {author} {\bibfnamefont
  {E.~D.}\ \bibnamefont {Hermes}}, \bibinfo {author} {\bibfnamefont {P.~C.}\
  \bibnamefont {Jennings}}, \bibinfo {author} {\bibfnamefont {P.~B.}\
  \bibnamefont {Jensen}}, \bibinfo {author} {\bibfnamefont {J.}~\bibnamefont
  {Kermode}}, \bibinfo {author} {\bibfnamefont {J.~R.}\ \bibnamefont
  {Kitchin}}, \bibinfo {author} {\bibfnamefont {E.~L.}\ \bibnamefont
  {Kolsbjerg}}, \bibinfo {author} {\bibfnamefont {J.}~\bibnamefont {Kubal}},
  \bibinfo {author} {\bibfnamefont {K.}~\bibnamefont {Kaasbjerg}}, \bibinfo
  {author} {\bibfnamefont {S.}~\bibnamefont {Lysgaard}}, \bibinfo {author}
  {\bibfnamefont {J.~B.}\ \bibnamefont {Maronsson}}, \bibinfo {author}
  {\bibfnamefont {T.}~\bibnamefont {Maxson}}, \bibinfo {author} {\bibfnamefont
  {T.}~\bibnamefont {Olsen}}, \bibinfo {author} {\bibfnamefont
  {L.}~\bibnamefont {Pastewka}}, \bibinfo {author} {\bibfnamefont
  {A.}~\bibnamefont {Peterson}}, \bibinfo {author} {\bibfnamefont
  {C.}~\bibnamefont {Rostgaard}}, \bibinfo {author} {\bibfnamefont
  {J.}~\bibnamefont {Schi{\o}tz}}, \bibinfo {author} {\bibfnamefont
  {O.}~\bibnamefont {Schütt}}, \bibinfo {author} {\bibfnamefont
  {M.}~\bibnamefont {Strange}}, \bibinfo {author} {\bibfnamefont {K.~S.}\
  \bibnamefont {Thygesen}}, \bibinfo {author} {\bibfnamefont {T.}~\bibnamefont
  {Vegge}}, \bibinfo {author} {\bibfnamefont {L.}~\bibnamefont {Vilhelmsen}},
  \bibinfo {author} {\bibfnamefont {M.}~\bibnamefont {Walter}}, \bibinfo
  {author} {\bibfnamefont {Z.}~\bibnamefont {Zeng}}, \ and\ \bibinfo {author}
  {\bibfnamefont {K.~W.}\ \bibnamefont {Jacobsen}},\ }\href@noop {} {\bibfield
  {journal} {\bibinfo  {journal} {J. Phys. Condens. Matter}\ }\textbf {\bibinfo
  {volume} {29}},\ \bibinfo {pages} {273002} (\bibinfo {year}
  {2017})}\BibitemShut {NoStop}%
\bibitem [{\citenamefont {Blum}\ \emph {et~al.}(2009)\citenamefont {Blum},
  \citenamefont {Gehrke}, \citenamefont {Hanke}, \citenamefont {Havu},
  \citenamefont {Havu}, \citenamefont {Ren}, \citenamefont {Reuter},\ and\
  \citenamefont {Scheffler}}]{Blum2009}%
  \BibitemOpen
  \bibfield  {author} {\bibinfo {author} {\bibfnamefont {V.}~\bibnamefont
  {Blum}}, \bibinfo {author} {\bibfnamefont {R.}~\bibnamefont {Gehrke}},
  \bibinfo {author} {\bibfnamefont {F.}~\bibnamefont {Hanke}}, \bibinfo
  {author} {\bibfnamefont {P.}~\bibnamefont {Havu}}, \bibinfo {author}
  {\bibfnamefont {V.}~\bibnamefont {Havu}}, \bibinfo {author} {\bibfnamefont
  {X.}~\bibnamefont {Ren}}, \bibinfo {author} {\bibfnamefont {K.}~\bibnamefont
  {Reuter}}, \ and\ \bibinfo {author} {\bibfnamefont {M.}~\bibnamefont
  {Scheffler}},\ }\href@noop {} {\bibfield  {journal} {\bibinfo  {journal}
  {Comput. Phys. Commun.}\ }\textbf {\bibinfo {volume} {180}},\ \bibinfo
  {pages} {2175 } (\bibinfo {year} {2009})}\BibitemShut {NoStop}%
\bibitem [{\citenamefont {Perdew}, \citenamefont {Burke},\ and\ \citenamefont
  {Ernzerhof}(1996)}]{Perdew1996}%
  \BibitemOpen
  \bibfield  {author} {\bibinfo {author} {\bibfnamefont {J.~P.}\ \bibnamefont
  {Perdew}}, \bibinfo {author} {\bibfnamefont {K.}~\bibnamefont {Burke}}, \
  and\ \bibinfo {author} {\bibfnamefont {M.}~\bibnamefont {Ernzerhof}},\
  }\href@noop {} {\bibfield  {journal} {\bibinfo  {journal} {Phys. Rev. Lett.}\
  }\textbf {\bibinfo {volume} {77}},\ \bibinfo {pages} {3865} (\bibinfo {year}
  {1996})}\BibitemShut {NoStop}%
\bibitem [{\citenamefont {Tkatchenko}\ and\ \citenamefont
  {Scheffler}(2009)}]{Tkatchenko2009}%
  \BibitemOpen
  \bibfield  {author} {\bibinfo {author} {\bibfnamefont {A.}~\bibnamefont
  {Tkatchenko}}\ and\ \bibinfo {author} {\bibfnamefont {M.}~\bibnamefont
  {Scheffler}},\ }\href@noop {} {\bibfield  {journal} {\bibinfo  {journal}
  {Phys. Rev. Lett.}\ }\textbf {\bibinfo {volume} {102}},\ \bibinfo {pages}
  {073005} (\bibinfo {year} {2009})}\BibitemShut {NoStop}%
\bibitem [{\citenamefont {Ruiz}\ \emph {et~al.}(2012)\citenamefont {Ruiz},
  \citenamefont {Liu}, \citenamefont {Zojer}, \citenamefont {Scheffler},\ and\
  \citenamefont {Tkatchenko}}]{Ruiz2012}%
  \BibitemOpen
  \bibfield  {author} {\bibinfo {author} {\bibfnamefont {V.~G.}\ \bibnamefont
  {Ruiz}}, \bibinfo {author} {\bibfnamefont {W.}~\bibnamefont {Liu}}, \bibinfo
  {author} {\bibfnamefont {E.}~\bibnamefont {Zojer}}, \bibinfo {author}
  {\bibfnamefont {M.}~\bibnamefont {Scheffler}}, \ and\ \bibinfo {author}
  {\bibfnamefont {A.}~\bibnamefont {Tkatchenko}},\ }\href@noop {} {\bibfield
  {journal} {\bibinfo  {journal} {Phys. Rev. Lett.}\ }\textbf {\bibinfo
  {volume} {108}},\ \bibinfo {pages} {146103} (\bibinfo {year}
  {2012})}\BibitemShut {NoStop}%
\bibitem [{Cu_(2011)}]{Cu_lattice_constant}%
  \BibitemOpen
  \enquote {\bibinfo {title} {Appendix {E}: Parameter tables of crystals},}\
  in\ \href {\doibase 10.1002/9783527633296.app5} {\emph {\bibinfo {booktitle}
  {Crystallography and Surface Structure}}}\ (\bibinfo  {publisher} {John Wiley
  \& Sons, Ltd},\ \bibinfo {year} {2011})\ pp.\ \bibinfo {pages}
  {265--266}\BibitemShut {NoStop}%
\bibitem [{\citenamefont {Olsen}\ \emph {et~al.}(2011)\citenamefont {Olsen},
  \citenamefont {Yan}, \citenamefont {Mortensen},\ and\ \citenamefont
  {Thygesen}}]{Olsen2011}%
  \BibitemOpen
  \bibfield  {author} {\bibinfo {author} {\bibfnamefont {T.}~\bibnamefont
  {Olsen}}, \bibinfo {author} {\bibfnamefont {J.}~\bibnamefont {Yan}}, \bibinfo
  {author} {\bibfnamefont {J.~J.}\ \bibnamefont {Mortensen}}, \ and\ \bibinfo
  {author} {\bibfnamefont {K.~S.}\ \bibnamefont {Thygesen}},\ }\href@noop {}
  {\bibfield  {journal} {\bibinfo  {journal} {Phys. Rev. Lett.}\ }\textbf
  {\bibinfo {volume} {107}},\ \bibinfo {pages} {156401} (\bibinfo {year}
  {2011})}\BibitemShut {NoStop}%
\bibitem [{\citenamefont {Andersen}, \citenamefont {Hornek\ae{}r},\ and\
  \citenamefont {Hammer}(2012)}]{Andersen2012}%
  \BibitemOpen
  \bibfield  {author} {\bibinfo {author} {\bibfnamefont {M.}~\bibnamefont
  {Andersen}}, \bibinfo {author} {\bibfnamefont {L.}~\bibnamefont
  {Hornek\ae{}r}}, \ and\ \bibinfo {author} {\bibfnamefont {B.}~\bibnamefont
  {Hammer}},\ }\href@noop {} {\bibfield  {journal} {\bibinfo  {journal} {Phys.
  Rev. B}\ }\textbf {\bibinfo {volume} {86}},\ \bibinfo {pages} {085405}
  (\bibinfo {year} {2012})}\BibitemShut {NoStop}%
\bibitem [{\citenamefont {Andersen}, \citenamefont {Cingolani},\ and\
  \citenamefont {Reuter}(2019)}]{Andersen2019b}%
  \BibitemOpen
  \bibfield  {author} {\bibinfo {author} {\bibfnamefont {M.}~\bibnamefont
  {Andersen}}, \bibinfo {author} {\bibfnamefont {J.~S.}\ \bibnamefont
  {Cingolani}}, \ and\ \bibinfo {author} {\bibfnamefont {K.}~\bibnamefont
  {Reuter}},\ }\href@noop {} {\bibfield  {journal} {\bibinfo  {journal} {J.
  Phys. Chem. C}\ }\textbf {\bibinfo {volume} {123}},\ \bibinfo {pages} {22299}
  (\bibinfo {year} {2019})}\BibitemShut {NoStop}%
\bibitem [{\citenamefont {Giannozzi}\ \emph {et~al.}(2017)\citenamefont
  {Giannozzi}, \citenamefont {Andreussi}, \citenamefont {Brumme}, \citenamefont
  {Bunau}, \citenamefont {Nardelli}, \citenamefont {Calandra}, \citenamefont
  {Car}, \citenamefont {Cavazzoni}, \citenamefont {Ceresoli}, \citenamefont
  {Cococcioni}, \citenamefont {Colonna}, \citenamefont {Carnimeo},
  \citenamefont {Corso}, \citenamefont {de~Gironcoli}, \citenamefont {Delugas},
  \citenamefont {Jr}, \citenamefont {Ferretti}, \citenamefont {Floris},
  \citenamefont {Fratesi}, \citenamefont {Fugallo}, \citenamefont {Gebauer},
  \citenamefont {Gerstmann}, \citenamefont {Giustino}, \citenamefont {Gorni},
  \citenamefont {Jia}, \citenamefont {Kawamura}, \citenamefont {Ko},
  \citenamefont {Kokalj}, \citenamefont {Küçükbenli}, \citenamefont
  {Lazzeri}, \citenamefont {Marsili}, \citenamefont {Marzari}, \citenamefont
  {Mauri}, \citenamefont {Nguyen}, \citenamefont {Nguyen}, \citenamefont {de-la
  Roza}, \citenamefont {Paulatto}, \citenamefont {Poncé}, \citenamefont
  {Rocca}, \citenamefont {Sabatini}, \citenamefont {Santra}, \citenamefont
  {Schlipf}, \citenamefont {Seitsonen}, \citenamefont {Smogunov}, \citenamefont
  {Timrov}, \citenamefont {Thonhauser}, \citenamefont {Umari}, \citenamefont
  {Vast}, \citenamefont {Wu},\ and\ \citenamefont {Baroni}}]{Giannozzi2017}%
  \BibitemOpen
  \bibfield  {author} {\bibinfo {author} {\bibfnamefont {P.}~\bibnamefont
  {Giannozzi}}, \bibinfo {author} {\bibfnamefont {O.}~\bibnamefont
  {Andreussi}}, \bibinfo {author} {\bibfnamefont {T.}~\bibnamefont {Brumme}},
  \bibinfo {author} {\bibfnamefont {O.}~\bibnamefont {Bunau}}, \bibinfo
  {author} {\bibfnamefont {M.~B.}\ \bibnamefont {Nardelli}}, \bibinfo {author}
  {\bibfnamefont {M.}~\bibnamefont {Calandra}}, \bibinfo {author}
  {\bibfnamefont {R.}~\bibnamefont {Car}}, \bibinfo {author} {\bibfnamefont
  {C.}~\bibnamefont {Cavazzoni}}, \bibinfo {author} {\bibfnamefont
  {D.}~\bibnamefont {Ceresoli}}, \bibinfo {author} {\bibfnamefont
  {M.}~\bibnamefont {Cococcioni}}, \bibinfo {author} {\bibfnamefont
  {N.}~\bibnamefont {Colonna}}, \bibinfo {author} {\bibfnamefont
  {I.}~\bibnamefont {Carnimeo}}, \bibinfo {author} {\bibfnamefont {A.~D.}\
  \bibnamefont {Corso}}, \bibinfo {author} {\bibfnamefont {S.}~\bibnamefont
  {de~Gironcoli}}, \bibinfo {author} {\bibfnamefont {P.}~\bibnamefont
  {Delugas}}, \bibinfo {author} {\bibfnamefont {R.~A.~D.}\ \bibnamefont {Jr}},
  \bibinfo {author} {\bibfnamefont {A.}~\bibnamefont {Ferretti}}, \bibinfo
  {author} {\bibfnamefont {A.}~\bibnamefont {Floris}}, \bibinfo {author}
  {\bibfnamefont {G.}~\bibnamefont {Fratesi}}, \bibinfo {author} {\bibfnamefont
  {G.}~\bibnamefont {Fugallo}}, \bibinfo {author} {\bibfnamefont
  {R.}~\bibnamefont {Gebauer}}, \bibinfo {author} {\bibfnamefont
  {U.}~\bibnamefont {Gerstmann}}, \bibinfo {author} {\bibfnamefont
  {F.}~\bibnamefont {Giustino}}, \bibinfo {author} {\bibfnamefont
  {T.}~\bibnamefont {Gorni}}, \bibinfo {author} {\bibfnamefont
  {J.}~\bibnamefont {Jia}}, \bibinfo {author} {\bibfnamefont {M.}~\bibnamefont
  {Kawamura}}, \bibinfo {author} {\bibfnamefont {H.-Y.}\ \bibnamefont {Ko}},
  \bibinfo {author} {\bibfnamefont {A.}~\bibnamefont {Kokalj}}, \bibinfo
  {author} {\bibfnamefont {E.}~\bibnamefont {Küçükbenli}}, \bibinfo {author}
  {\bibfnamefont {M.}~\bibnamefont {Lazzeri}}, \bibinfo {author} {\bibfnamefont
  {M.}~\bibnamefont {Marsili}}, \bibinfo {author} {\bibfnamefont
  {N.}~\bibnamefont {Marzari}}, \bibinfo {author} {\bibfnamefont
  {F.}~\bibnamefont {Mauri}}, \bibinfo {author} {\bibfnamefont {N.~L.}\
  \bibnamefont {Nguyen}}, \bibinfo {author} {\bibfnamefont {H.-V.}\
  \bibnamefont {Nguyen}}, \bibinfo {author} {\bibfnamefont {A.~O.}\
  \bibnamefont {de-la Roza}}, \bibinfo {author} {\bibfnamefont
  {L.}~\bibnamefont {Paulatto}}, \bibinfo {author} {\bibfnamefont
  {S.}~\bibnamefont {Poncé}}, \bibinfo {author} {\bibfnamefont
  {D.}~\bibnamefont {Rocca}}, \bibinfo {author} {\bibfnamefont
  {R.}~\bibnamefont {Sabatini}}, \bibinfo {author} {\bibfnamefont
  {B.}~\bibnamefont {Santra}}, \bibinfo {author} {\bibfnamefont
  {M.}~\bibnamefont {Schlipf}}, \bibinfo {author} {\bibfnamefont {A.~P.}\
  \bibnamefont {Seitsonen}}, \bibinfo {author} {\bibfnamefont {A.}~\bibnamefont
  {Smogunov}}, \bibinfo {author} {\bibfnamefont {I.}~\bibnamefont {Timrov}},
  \bibinfo {author} {\bibfnamefont {T.}~\bibnamefont {Thonhauser}}, \bibinfo
  {author} {\bibfnamefont {P.}~\bibnamefont {Umari}}, \bibinfo {author}
  {\bibfnamefont {N.}~\bibnamefont {Vast}}, \bibinfo {author} {\bibfnamefont
  {X.}~\bibnamefont {Wu}}, \ and\ \bibinfo {author} {\bibfnamefont
  {S.}~\bibnamefont {Baroni}},\ }\href@noop {} {\bibfield  {journal} {\bibinfo
  {journal} {J. Phys. Condens. Matter}\ }\textbf {\bibinfo {volume} {29}},\
  \bibinfo {pages} {465901} (\bibinfo {year} {2017})}\BibitemShut {NoStop}%
\bibitem [{\citenamefont {Grimme}\ \emph {et~al.}(2010)\citenamefont {Grimme},
  \citenamefont {Antony}, \citenamefont {Ehrlich},\ and\ \citenamefont
  {Krieg}}]{Grimme2010}%
  \BibitemOpen
  \bibfield  {author} {\bibinfo {author} {\bibfnamefont {S.}~\bibnamefont
  {Grimme}}, \bibinfo {author} {\bibfnamefont {J.}~\bibnamefont {Antony}},
  \bibinfo {author} {\bibfnamefont {S.}~\bibnamefont {Ehrlich}}, \ and\
  \bibinfo {author} {\bibfnamefont {H.}~\bibnamefont {Krieg}},\ }\href@noop {}
  {\bibfield  {journal} {\bibinfo  {journal} {J. Chem. Phys.}\ }\textbf
  {\bibinfo {volume} {132}},\ \bibinfo {pages} {154104} (\bibinfo {year}
  {2010})}\BibitemShut {NoStop}%
\bibitem [{\citenamefont {Bengtsson}(1999)}]{Bengtsson1999}%
  \BibitemOpen
  \bibfield  {author} {\bibinfo {author} {\bibfnamefont {L.}~\bibnamefont
  {Bengtsson}},\ }\href@noop {} {\bibfield  {journal} {\bibinfo  {journal}
  {Phys. Rev. B}\ }\textbf {\bibinfo {volume} {59}},\ \bibinfo {pages} {12301}
  (\bibinfo {year} {1999})}\BibitemShut {NoStop}%
\bibitem [{\citenamefont {Zhong}\ \emph {et~al.}(2016)\citenamefont {Zhong},
  \citenamefont {Li}, \citenamefont {Li}, \citenamefont {Lu}, \citenamefont
  {Du}, \citenamefont {Gan}, \citenamefont {Xu}, \citenamefont {Chiang},\ and\
  \citenamefont {Kang}}]{Zhong2016}%
  \BibitemOpen
  \bibfield  {author} {\bibinfo {author} {\bibfnamefont {L.}~\bibnamefont
  {Zhong}}, \bibinfo {author} {\bibfnamefont {J.}~\bibnamefont {Li}}, \bibinfo
  {author} {\bibfnamefont {Y.}~\bibnamefont {Li}}, \bibinfo {author}
  {\bibfnamefont {H.}~\bibnamefont {Lu}}, \bibinfo {author} {\bibfnamefont
  {H.}~\bibnamefont {Du}}, \bibinfo {author} {\bibfnamefont {L.}~\bibnamefont
  {Gan}}, \bibinfo {author} {\bibfnamefont {C.}~\bibnamefont {Xu}}, \bibinfo
  {author} {\bibfnamefont {S.~W.}\ \bibnamefont {Chiang}}, \ and\ \bibinfo
  {author} {\bibfnamefont {F.}~\bibnamefont {Kang}},\ }\href@noop {} {\bibfield
   {journal} {\bibinfo  {journal} {J. Phys. Chem. C}\ }\textbf {\bibinfo
  {volume} {120}},\ \bibinfo {pages} {23239} (\bibinfo {year}
  {2016})}\BibitemShut {NoStop}%
\bibitem [{\citenamefont {Didar}, \citenamefont {Khosravian},\ and\
  \citenamefont {Balbuena}(2018)}]{Didar2018}%
  \BibitemOpen
  \bibfield  {author} {\bibinfo {author} {\bibfnamefont {B.~R.}\ \bibnamefont
  {Didar}}, \bibinfo {author} {\bibfnamefont {H.}~\bibnamefont {Khosravian}}, \
  and\ \bibinfo {author} {\bibfnamefont {P.~B.}\ \bibnamefont {Balbuena}},\
  }\href@noop {} {\bibfield  {journal} {\bibinfo  {journal} {RSC Adv.}\
  }\textbf {\bibinfo {volume} {8}},\ \bibinfo {pages} {27825} (\bibinfo {year}
  {2018})}\BibitemShut {NoStop}%
\bibitem [{\citenamefont {H\"akkinen}\ and\ \citenamefont
  {Manninen}(1992)}]{Hakkinen1992}%
  \BibitemOpen
  \bibfield  {author} {\bibinfo {author} {\bibfnamefont {H.}~\bibnamefont
  {H\"akkinen}}\ and\ \bibinfo {author} {\bibfnamefont {M.}~\bibnamefont
  {Manninen}},\ }\href@noop {} {\bibfield  {journal} {\bibinfo  {journal}
  {Phys. Rev. B}\ }\textbf {\bibinfo {volume} {46}},\ \bibinfo {pages} {1725}
  (\bibinfo {year} {1992})}\BibitemShut {NoStop}%
\bibitem [{\citenamefont {Wang}\ \emph {et~al.}(2015)\citenamefont {Wang},
  \citenamefont {Weinberg}, \citenamefont {Zhang}, \citenamefont {Lunkenbein},
  \citenamefont {Klein-Hoffmann}, \citenamefont {Kurnatowska}, \citenamefont
  {Plodinec}, \citenamefont {Li}, \citenamefont {Chi}, \citenamefont
  {Schloegl},\ and\ \citenamefont {Willinger}}]{Wang2015b}%
  \BibitemOpen
  \bibfield  {author} {\bibinfo {author} {\bibfnamefont {Z.-J.}\ \bibnamefont
  {Wang}}, \bibinfo {author} {\bibfnamefont {G.}~\bibnamefont {Weinberg}},
  \bibinfo {author} {\bibfnamefont {Q.}~\bibnamefont {Zhang}}, \bibinfo
  {author} {\bibfnamefont {T.}~\bibnamefont {Lunkenbein}}, \bibinfo {author}
  {\bibfnamefont {A.}~\bibnamefont {Klein-Hoffmann}}, \bibinfo {author}
  {\bibfnamefont {M.}~\bibnamefont {Kurnatowska}}, \bibinfo {author}
  {\bibfnamefont {M.}~\bibnamefont {Plodinec}}, \bibinfo {author}
  {\bibfnamefont {Q.}~\bibnamefont {Li}}, \bibinfo {author} {\bibfnamefont
  {L.}~\bibnamefont {Chi}}, \bibinfo {author} {\bibfnamefont {R.}~\bibnamefont
  {Schloegl}}, \ and\ \bibinfo {author} {\bibfnamefont {M.-G.}\ \bibnamefont
  {Willinger}},\ }\href@noop {} {\bibfield  {journal} {\bibinfo  {journal} {ACS
  Nano}\ }\textbf {\bibinfo {volume} {9}},\ \bibinfo {pages} {1506} (\bibinfo
  {year} {2015})}\BibitemShut {NoStop}%
\bibitem [{\citenamefont {Stock}(1980)}]{Stock1980}%
  \BibitemOpen
  \bibfield  {author} {\bibinfo {author} {\bibfnamefont {K.}~\bibnamefont
  {Stock}},\ }\href@noop {} {\bibfield  {journal} {\bibinfo  {journal} {Surf.
  Sci.}\ }\textbf {\bibinfo {volume} {91}},\ \bibinfo {pages} {655 } (\bibinfo
  {year} {1980})}\BibitemShut {NoStop}%
\bibitem [{\citenamefont {Gromov}\ \emph {et~al.}(2007)\citenamefont {Gromov},
  \citenamefont {Gavrilov}, \citenamefont {Redichev},\ and\ \citenamefont
  {Ammosov}}]{Gromov2007}%
  \BibitemOpen
  \bibfield  {author} {\bibinfo {author} {\bibfnamefont {D.~G.}\ \bibnamefont
  {Gromov}}, \bibinfo {author} {\bibfnamefont {S.~A.}\ \bibnamefont
  {Gavrilov}}, \bibinfo {author} {\bibfnamefont {E.~N.}\ \bibnamefont
  {Redichev}}, \ and\ \bibinfo {author} {\bibfnamefont {R.~M.}\ \bibnamefont
  {Ammosov}},\ }\href@noop {} {\bibfield  {journal} {\bibinfo  {journal} {Phys.
  Solid State}\ }\textbf {\bibinfo {volume} {49}},\ \bibinfo {pages} {178}
  (\bibinfo {year} {2007})}\BibitemShut {NoStop}%
\bibitem [{\citenamefont {Zeng}\ \emph {et~al.}(2014)\citenamefont {Zeng},
  \citenamefont {Tan}, \citenamefont {Wang}, \citenamefont {Chen},
  \citenamefont {Rümmeli},\ and\ \citenamefont {Fu}}]{Zeng2014}%
  \BibitemOpen
  \bibfield  {author} {\bibinfo {author} {\bibfnamefont {M.}~\bibnamefont
  {Zeng}}, \bibinfo {author} {\bibfnamefont {L.}~\bibnamefont {Tan}}, \bibinfo
  {author} {\bibfnamefont {J.}~\bibnamefont {Wang}}, \bibinfo {author}
  {\bibfnamefont {L.}~\bibnamefont {Chen}}, \bibinfo {author} {\bibfnamefont
  {M.~H.}\ \bibnamefont {Rümmeli}}, \ and\ \bibinfo {author} {\bibfnamefont
  {L.}~\bibnamefont {Fu}},\ }\href@noop {} {\bibfield  {journal} {\bibinfo
  {journal} {Chem. Mater.}\ }\textbf {\bibinfo {volume} {26}},\ \bibinfo
  {pages} {3637} (\bibinfo {year} {2014})}\BibitemShut {NoStop}%
\bibitem [{\citenamefont {Bartelt}\ and\ \citenamefont
  {McCarty}(2012)}]{Bartelt2012}%
  \BibitemOpen
  \bibfield  {author} {\bibinfo {author} {\bibfnamefont {N.}~\bibnamefont
  {Bartelt}}\ and\ \bibinfo {author} {\bibfnamefont {K.}~\bibnamefont
  {McCarty}},\ }\href@noop {} {\bibfield  {journal} {\bibinfo  {journal} {MRS
  Bull.}\ }\textbf {\bibinfo {volume} {37}},\ \bibinfo {pages} {1158–1165}
  (\bibinfo {year} {2012})}\BibitemShut {NoStop}%
\bibitem [{\citenamefont {{J\"{u}lich Supercomputing Centre}}(2019)}]{JUWELS}%
  \BibitemOpen
  \bibfield  {author} {\bibinfo {author} {\bibnamefont {{J\"{u}lich
  Supercomputing Centre}}},\ }\href@noop {} {\bibfield  {journal} {\bibinfo
  {journal} {Journal of large-scale research facilities}\ }\textbf {\bibinfo
  {volume} {5}} (\bibinfo {year} {2019})}\BibitemShut {NoStop}%
\end{thebibliography}%

\end{document}